\documentclass[aps,prb,twocolumn,showpacs,amssymb,amsfonts]{revtex4-1}

\usepackage[T1]{fontenc}
\usepackage[latin9]{inputenc}
\usepackage{amsmath}
\usepackage{dsfont}
\usepackage{amstext}
\usepackage{amsthm}
\usepackage{tocvsec2}
\usepackage{amssymb}
\usepackage{amsbsy}
\usepackage{amsthm}
\usepackage{epsfig}
\usepackage{framed}
\usepackage{graphicx}
\usepackage{bbm}
\usepackage{hyperref}
\usepackage{color}
\usepackage{multirow}

\begin{document}
\title{Competing phases and topological excitations of spin-one pyrochlore antiferromagnets}
\author{Fei-Ye Li}
\affiliation{Department of Physics, 
Center for Field Theory and Particle Physics, 
State Key Laboratory of Surface Physics,
Fudan University, Shanghai 200433, China}
\author{Gang Chen}
\email{gangchen.physics@gmail.com}
\affiliation{Department of Physics, 
Center for Field Theory and Particle Physics, 
State Key Laboratory of Surface Physics,
Fudan University, Shanghai 200433, China}
\affiliation{Collaborative Innovation Center of Advanced 
Microstructures, Nanjing, 210093, China}

\date{\today}

\begin{abstract}
Most works on pyrochlore magnets deal with the interacting spin-1/2 local moments. 
We here study the spin-one local moments on the pyrochlore lattice, and propose 
a generic interacting spin model on a pyrochlore lattice. 
Our spin model includes the antiferromagnetic Heisenberg 
interaction, the Dzyaloshinskii-Moriya interaction and the single-ion spin
anisotropy. We develop a flavor wave theory and combine with a mean-field approach 
to study the global phase diagram of this model and establish the relation between
different phases in the phase diagram. We find the regime of the quantum paramagnetic 
phase where a degenerate line of the magnetic excitations emerges in the momentum 
space. We further predict the critical properties of the transition out of the 
quantum paramagnet to the proximate orders. The presence of quantum order by 
disorder in the parts of the ordered phases is then suggested. We point out 
the existence of degenerate and topological excitations in various phases. We 
discuss the relevance with fluoride pyrochlore material NaCaNi$_2$F$_7$ and 
explain the role of the spin-orbit coupling and the magnetic structures of 
the Ru-based pyrochlore A$_2$Ru$_2$O$_7$ and the Mo-based pyrochlore 
A$_2$Mo$_2$O$_7$.
\end{abstract}

\maketitle

\section{Introduction}

Recently, there is a growing interest and effort in the frustrated magnetic 
systems with spin-one local moments, and interesting quantum phases and 
unconventional excitations have been predicted for frustrated spin-one 
systems~\cite{Haldane1,Haldane2,AKLT,XChenXGWen,WangSenthil,Chen2017,Savary2017,Oleg2017}. 
In particular, a chiral liquid phase with a finite vector chirality order has 
been obtained for the spin-one triangular lattice magnet~\cite{Oleg2017}, 
Haldane phase like symmetry-protected topological phases have been suggested 
for three-dimensional spin-one systems~\cite{WangSenthil,mcqueen2017}, spin liquid related 
physics and phenomenology has been explored for the layered triangular 
material Ba$_3$NiSb$_2$O$_9$~\cite{Cheng2011,Serbyn2011,Samuel2012,Cenke2012,Chen2012,Hwang2013,Quilliam2016}, 
and exotic excitations with degenerate band minima were established for 
the spin-one diamond lattice antiferromagnet~\cite{Chen2017,Trebst2017}. 
In this work, we turn our attention to the spin-one pyrochlore lattice antiferromagnet. 

Pyrochlore antiferromagnet~\cite{RevModPhys.82.53} is a stereotype of spin systems with 
geometrical frustration and potential quantum phases. In last decade or so,
most efforts in the field were devoted to the rare-earth pyrochlore 
magnets where the relevant degrees of freedom are certain spin-orbital-entangled 
effective spin-1/2 local moments~\cite{RevModPhys.82.53,Gingras2001,
PhysRevLett.87.067203,Castelnovo12008,PhysRevLett.98.157204,Gingras2014,
BalentsSavary,PhysRevLett.105.047201,Savary12,Sungbin2012,SavaryPRB,
PhysRevLett.87.067203,PhysRevB.65.054410,PhysRevLett.87.047205,Gingras2014,
Ross2009,Huang2014,PhysRevB.94.205107,PhysRevLett.108.247210,PhysRevB.95.041106,
PhysRevB.95.094422,Chen2015,SavaryPRB,PhysRevLett.109.017201,Yasui2002,
PhysRevB.64.224416,PhysRevB.90.214430,Chang2012,Kimura2012,Lhotel2014,Chang2014,Yasui2003,
Ross11,Shannon12,Goswami2016,PhysRevB.95.094407,PhysRevLett.118.107206,
PhysRevB.92.054432,PhysRevLett.113.197202,fu2017fingerprints,PhysRevB.86.075154,
PhysRevLett.115.267208,PhysRevLett.109.097205,PhysRevLett.107.207207,PhysRevLett.115.097202,
Shannon2017,GangChen2017,PhysRevLett.118.087203,PhysRevB.96.195127}. Due to the geometrical 
frustration and the bond-dependent anisotropic spin 
interaction~\cite{PhysRevLett.87.067203,Gingras2001,
PhysRevLett.105.047201,PhysRevB.78.094418,1742-6596-320-1-012065}, 
interesting magnetic phases and phenomena, quantum spin ice and $U(1)$ quantum spin 
liquid for example, have been 
proposed and explored~\cite{PhysRevLett.98.157204,Savary12,Sungbin2012,PhysRevLett.105.047201}. 
This field is fertilized by the existence of the abundant rare-earth pyrochlore 
magnets with different magnetic ions. Recently, a new 
family of fluoride pyrochlore systems with the transition metal ions Fe$^{2+}$, 
Co$^{2+}$, Ni$^{2+}$ and Mn$^{2+}$ has been synthesized~\cite{PhysRevB.92.014406,
PhysRevB.89.214401,PhysRevB.95.144414,0953-8984-29-4-045801}. 
Unlike the rare-earth $4f$ electrons whose interactions are usually quite small, 
these new systems, consisting of transition metal ions, have much stronger spin 
interactions. Moreover, spin-orbit coupling is less important in these systems, 
although spin-orbit coupling sometimes becomes active and modifies the local 
moment structure if there exists a partially filled $t_{2g}$ shell for the magnetic 
ions~\cite{witczak2014correlated}. 

\begin{figure}[b]
\includegraphics[width=8.5cm]{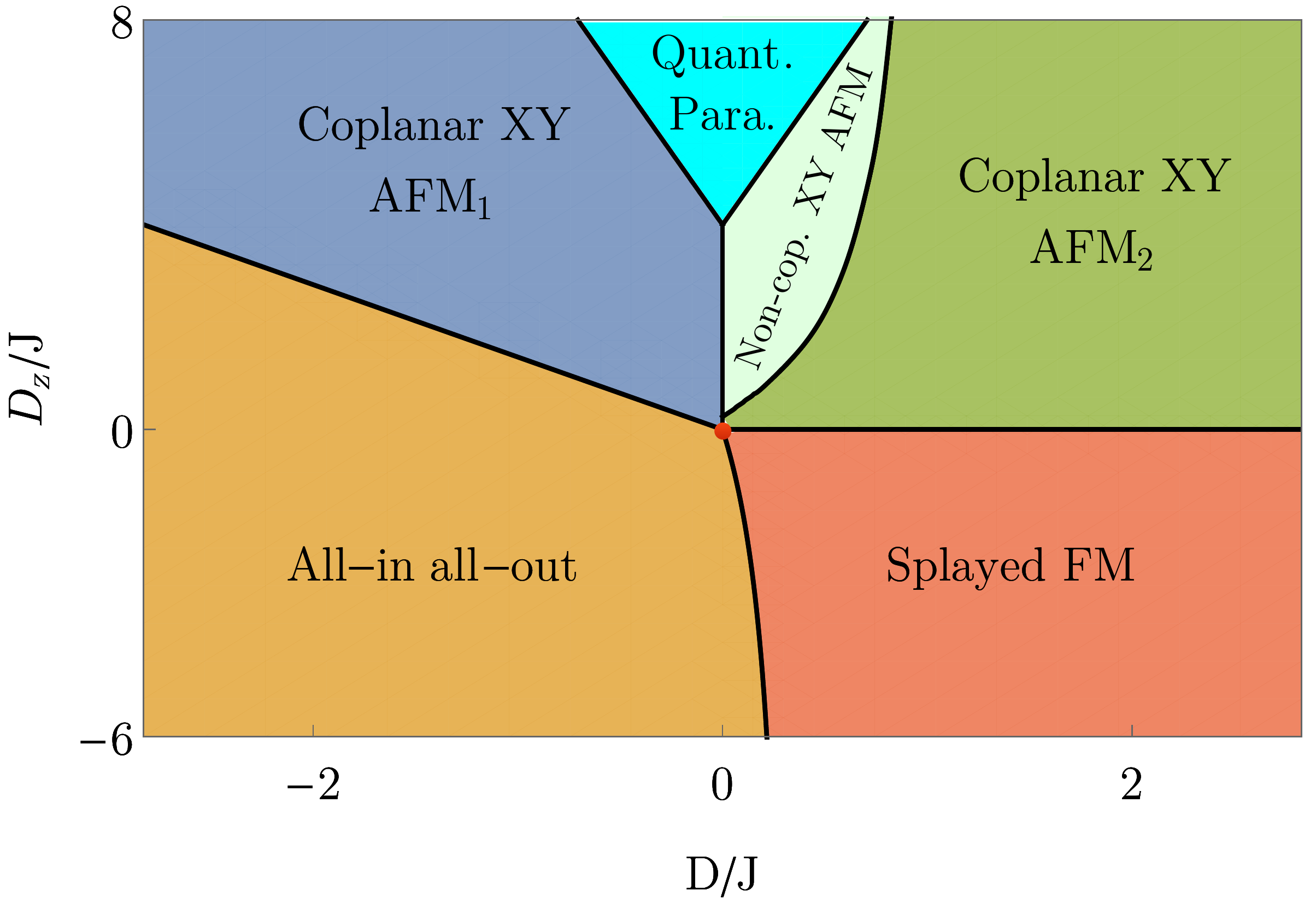}
\caption{The phase diagram of our generic spin model for the spin-1 pyrochlore system.  
Here, the Heisenberg exchange $J$ is set to be antiferromagnetic 
with ${J>0}$. ``Quant Para'' refers to the quantum paramagnetic phase. 
The details of the ordered phases are explained in the main text. 
The (red) dot is the Heisenberg point of the model. 
A similar phase diagram with the ferromagnetic Heisenberg exchange 
is found in the Appendix~\ref{app5}.}
\label{fig1}
\end{figure}

Just like the fundamental distinction between the half-integer and the 
integer spin moments for one dimensional spin chains that was pointed out
by F.D.M. Haldane~\cite{Haldane1,Haldane2}, the physical properties
of the half-integer spin and the integer spin moments on the pyrochlore 
lattice are expected to be quite different. In fact, for the rare-earth 
pyrochlore magnets, such a distinction has already been manifested in 
the Kramers doublet system and the non-Kramers doublet system where 
the non-Kramers doublet originates from integer spin and supports 
magnetic quadrupolar order~\cite{PhysRevLett.105.047201,Sungbin2012,PhysRevB.94.205107}. 
Since most works in this field are dealing with effective
spin-1/2 pyrochlores, it is valuable to consider the physics
of the spin-1 pyrochlores. 

Among the existing fluoride pyrochlores,
Co$^{2+}$ and Mn$^{2+}$ have half-integer spin moments while 
Ni$^{2+}$ and Fe$^{2+}$ have integer spin moments~\cite{PhysRevB.92.014406,
PhysRevB.89.214401,PhysRevB.95.144414,0953-8984-29-4-045801}.
From the conventional wisdom, when the spin moment is large, 
the system tends to behave more classically. For geometrically frustrated
systems, however, the spin-one local moments may occasionally give
rise to quantum phenomena. Indeed, in the Ni-based fluoride pyrochlore 
NaCaNi$_2$F$_7$, spin-ordering-related features were not found 
in the thermodynamic measurement down to the spin glassy transition 
at $3.6$K that is attributed to the possible bond randomness, 
although the system has the Curie-Weiss temperature $-129$K~\cite{PhysRevB.92.014406}. 
Apart from this new material, the spin-one pyrochlores have already
been suggested for the Ru-based pyrochlore A$_2$Ru$_2$O$_7$ and
the Mo-based pyrochlore A$_2$Mo$_2$O$_7$, despite the fact that 
the stronger spin-orbit coupling of the 
$4d$ electrons may be more important in these two systems. 
Partly motivated by these experiments and more broadly about 
the physics of the spin-one moments, in this paper, 
we study the generic spin model and the magnetic properties 
of the spin-one local moments on the pyrochlore lattice.

We point out that, in addition to the Heisenberg model that 
is usually assumed for the $3d$ transition metal ions 
and sometimes for the $4d$ transition metal ions, 
there exist the on-site single-ion spin anisotropy 
and the antisymmetric Dyzaloshinskii-Moriya interaction. 
Our phase diagram is summarized in Fig.~\ref{fig1}. 
In our approach, we start from the quantum paramagnetic 
ground state in the strong single-ion spin anisotropic limit 
and explore the instability of this quantum state as the 
Heisenberg exchange and the Dyzaloshinskii-Moriya interaction
are switched on. Mostly relying on a flavor wave theory, 
we access the phase transitions out of this quantum paramagnetic 
state and explore the properties of criticalities. Inside the 
ordered phases, we implement the usual mean-field theory 
and establish the phase diagram on the ordered side. 
We further identify the region on the ordered side where 
there exist continuous degeneracies of the ground state 
manifold at the mean-field level. The quantum fluctuation 
is studied and lifts the continuous degeneracies. 
The magnetic excitations in different phases
are also discussed.

The following parts of the paper are organized as follows. 
In Sec.~\ref{sec2}, we introduce the model Hamiltonian. 
In Sec.~\ref{sec3}, we use the flavor wave theory and 
study the magnetic excitation 
and the instability of the quantum paramagnetic phase.
In Sec.~\ref{sec4}, we focus on the ordered side and 
study the magnetic properties of the magnetic orders. 
Finally in Sec.~\ref{sec5}, we summarize the theoretical 
prediction and the physical properties of the phase diagram, 
discuss the materials' relevance, and make an extension
to spin-3/2 pyrochlores.

\section{Model Hamiltonian}
\label{sec2}

\begin{figure}[b]
\includegraphics[width=7.8cm]{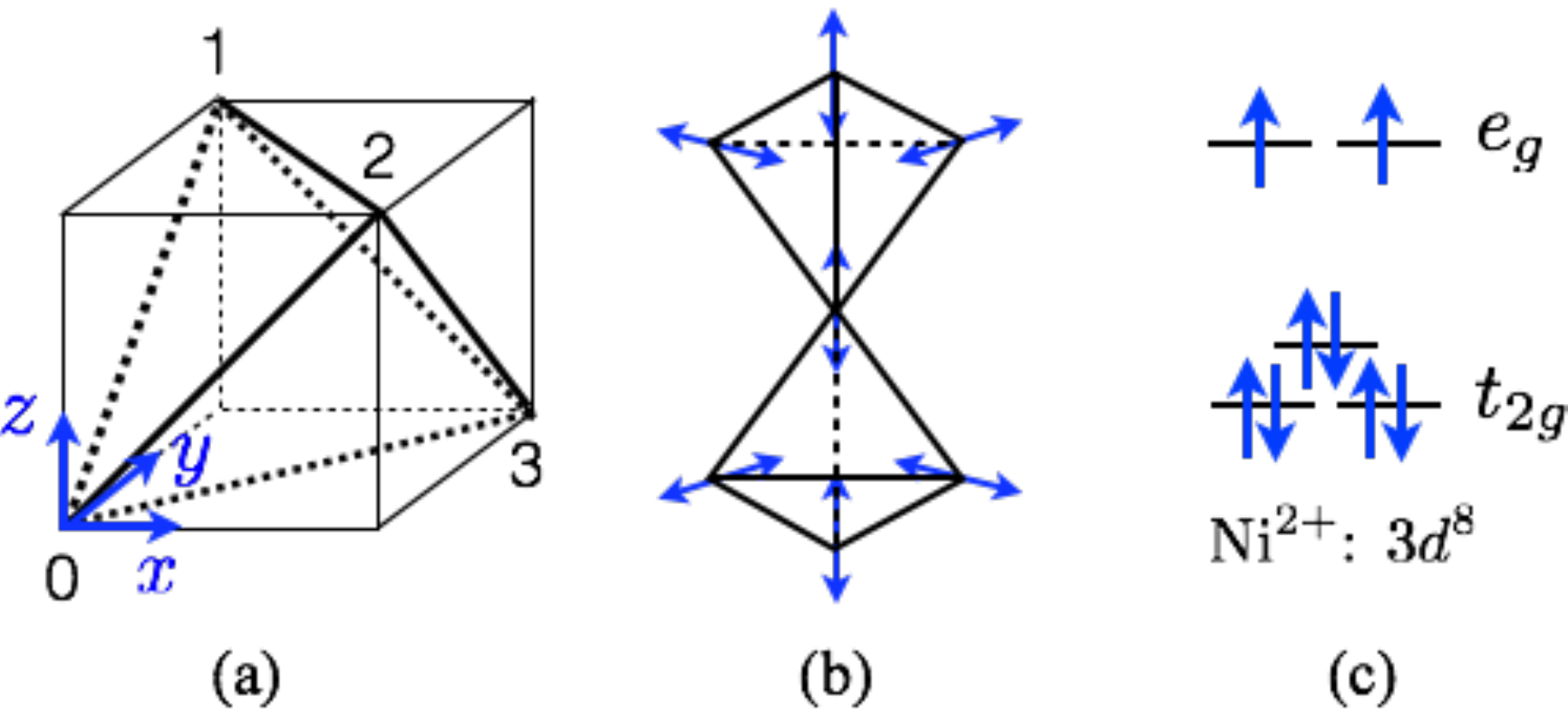}
\caption{(a) The four sublattices and the unit cell of the 
pyrochlore lattice. (b) The (blue) arrows define the local $z$ or 
$\langle 111 \rangle$ axis. 
(c) The electron configuration of the Ni$^{2+}$ ion in NaCaNi$_2$F$_7$. 
While the $e_g$ orbitals remain degenerate under the D$_{3d}$ point group, 
the $t_{2g}$ orbitals would be broken into $a_{1g}$ and two-fold 
degenerate $e_{g}'$ orbitals. The relative energies of $a_{1g}$ and 
$e_g'$ orbitals are unknown, and we place $a_{1g}$ at a higher energy in
the figure. The ${S=1}$ nature of the Ni$^{2+}$ local moment holds for 
either distribution of the $a_{1g}$ and $e_g'$ orbitals.}
\label{fig2}
\end{figure}

We start from the local moment physics of the Ni$^{2+}$ ion in NaCaNi$_2$F$_7$.
Although the starting point here is specific to NaCaNi$_2$F$_7$, the physical 
model itself applies broadly to other spin-one pyrochlore systems, and we merely
present the model through the specific case of NaCaNi$_2$F$_7$. 
The Ni$^{2+}$ ion has a $3d^8$ electron configuration. 
In the octahedral crystal field environment of NaCaNi$_2$F$_7$, 
the six electrons occupy the lower $t_{2g}$ orbitals, and the 
remaining two electrons occupy the upper $e_g$ orbitals and form 
a spin ${S=1}$ local moment. There is no orbital degeneracy here. 
We propose the following spin model for the interaction between 
the local moments. The minimal spin Hamiltonian is given as~\cite{1742-6596-320-1-012065},
\begin{eqnarray}
H & = & \sum_{\langle ij \rangle} \big[ J {\boldsymbol S}_i \cdot {\boldsymbol S}_j 
+ {\boldsymbol D}_{ij} \cdot ({\boldsymbol S}_i \times {\boldsymbol S}_j ) \big] 
\nonumber \\
&& + \sum_i D_z ({\boldsymbol S}_i \cdot \hat{z}_i)^2 ,
\label{eq1}
\end{eqnarray}
where ${\boldsymbol D}_{ij}$ is the bond-dependent vector that 
defines the antisymmetric Dzyaloshinskii-Moriya interaction~\cite{PhysRevB.71.094420}. 
For the 01 bond in Fig.~\ref{fig2}a, 
we have 
\begin{equation}
{\boldsymbol D}_{01} = (0,  \frac{D}{\sqrt{2}},  - \frac{D}{\sqrt{2}} ),
\end{equation}
and ${\boldsymbol D}_{ij}$'s on other bonds are readily obtained from      
the lattice symmmetry. The $D_z$ term is the single-ion spin anisotropy 
allowed by the D$_{3d}$ point group symmetry of the pyrochlore lattice, 
and $\hat{z}_i$ is the local $\langle 111 \rangle$ axis that 
is defined locally for each pyrochlore sublattice. Even though the 
Dzyaloshinskii-Moriya interaction arises from the first order effect 
of the spin-orbit coupling and the single-ion spin anisotropy arises 
from the second order effect of the spin-orbit coupling, it does not 
necessarily indicate the single-ion anisotropy is weaker than 
the Dzyaloshinskii-Moriya interaction. In fact, ignoring the
effect from Hund's coupling, one has the following results~\cite{TMO} 
\begin{eqnarray}
|D_{ij}|/J & \sim & \mathcal{O}(\lambda/ \Delta ) ,   
 \label{Eqdm} 
\\
|D_z|/\Delta & \sim & \mathcal{O}(\lambda^2/\Delta^2), 
\label{Eqion}
\end{eqnarray}
where $\lambda$ is the spin-orbit coupling and $\Delta$ 
is the crystal electric field splitting between the $t_{2g}$ 
and the $e_g$ manifolds and can be much larger than the 
superexchange interaction $J$. As a result, whether $\lambda$ 
appears as the linear order or as the second order cannot be 
used to argue for the relative magnitudes of $|D_{ij}|$ and $D_z$. 
We include both couplings in our model Hamiltonian. 
We have neglected the pseudo-dipolar interactions, as they are 
subleading compared to the Dzyaloshinskii-Moriya interaction
for the $3d$ transition metal ions without any orbital degeneracy~\cite{PhysRev.120.91}.
The pseudo-dipolar interactions, however, may become important
for the $4d$ transition metal ions.

\section{Flavor wave theory for quantum paramagnet}
\label{sec3}

Our minimal model contains three different interactions. The quantum ground 
state of the Heisenberg model is one of the hardest problems in quantum magnetism, 
so it is not so profitable to start from there. Instead, we start from the 
strong single-ion spin anisotropy limit with ${D_z >0}$ where the ground   
state is a simple product state of the quantum paramagnet with 
\begin{equation}
|\text{quantum paramagnet} \rangle  
= \prod_i | { S^{z}_i\equiv {\boldsymbol S}_i \cdot \hat{z}_i = 0 } \rangle. 
\end{equation}
This state is impossible for the half-integer spin local moments as there is always
Kramers' degeneracy. From this well-understood limit, we turn on the exchange 
interaction and study the evolution of the magnetic excitation and the instability. 

For our convenience, we first rewrite the spin Hamiltonian in the local  
coordinate basis since the single-ion anisotropy is defined locally.     
Under the local coordinate systems that are defined in the Appendix~\ref{app2}, our 
spin model reduces to~\cite{1742-6596-320-1-012065}
\begin{eqnarray}
H & = & \sum_{\langle ij \rangle} \big[ J_{zz} S_i^z S_j^z + J_{\pm} (S^+_i S^-_j + h.c. )
+ J_{\pm\pm} (\gamma_{ij}^{} S^+_i S^+_j 
\nonumber \\
&& \quad + \gamma_{ij}^{\ast} S^-_i S^-_j  ) 
   + J_{z\pm} (\xi_{ij}^{} S_i^z  S_j^+  + \xi_{ij}^{} S_i^+  S_j^z + h.c.) \big]
\nonumber \\
&& \quad 
+ \sum_i D_z (S_i^z)^2,
\label{eq4}
\end{eqnarray} 
where these spin operators, $S^z_i, S^+_i, S^-_i$, are defined in the 
local coordinate system for each sublattice. 
Note the exchange part of the model has the general form as 
the one for the Kramers doublet on the pyrochlore lattice, 
and the bond dependent phase variables $\gamma_{ij}$ and
$\xi_{ij}$ where $\gamma_{ij}$ takes $1, e^{i2\pi/3}, e^{-i2\pi/3}$
for the bonds on different planes and ${\xi_{ij} = -\gamma_{ij}^{\ast}}$.      
The relation between the couplings in the above equation
and the couplings in Eq.~\eqref{eq1} is listed in Appendix~\ref{app2}. 
In the following, we will focus our analysis on this form of the
model.

\subsection{Flavor wave representation}

This quantum paramagnet has no long-range magnetic order, and the 
conventional Holstein-Primarkoff spin-wave theory cannot be directly 
applied at all. For our purpose, we invoke so-called flavor wave theory, 
that was first developed in Ref.~\onlinecite{FCZhang} 
for the $SU(4)$ spin-orbital model~\cite{PhysRevLett.81.3527}, 
and properly adjust the formulation to our case. We define the states 
in the Hilbert space as 
\begin{eqnarray}
|m \rangle_i^{} \equiv | { S^{z}_i  = m } \rangle,
\end{eqnarray}
where ${m=0,\pm 1}$, and the elementary operator is then given as  
$S^n_m (i) \equiv |m\rangle_i^{} \langle n|_i^{}$. For the quantum 
paramagnet, we introduce the following flavor-wave representation, 
\begin{eqnarray}
S^0_0 (i) &=& 1 - a^\dagger_1 (i) a^{}_1 (i) -a^\dagger_{\bar{1}} (i) 
a^{\phantom\dagger}_{\bar{1}} (i) ,
\label{eq5}
\\
S^0_1 (i) &=& a^\dagger_1 (i)  
              \big[ 1 - a^\dagger_1 (i) a^{}_1 (i) -a^\dagger_{\bar{1}} (i) 
              a^{\phantom\dagger}_{\bar{1}} (i) \big]^{\frac{1}{2}}, 
\label{eq6}             
\\
S^0_{\bar{1}} (i) &=& a^\dagger_{\bar{1}} (i)  
              \big[ 1 - a^\dagger_1 (i) a^{}_1 (i)-a^\dagger_{\bar{1}} (i) 
              a^{\phantom\dagger}_{\bar{1}} (i) \big]^{\frac{1}{2}},
\label{eq7}              
\\
S^1_{\bar{1}} (i) &=& a^\dagger_{\bar{1}} (i)  a^{\phantom\dagger}_{1} (i) , 
\label{eq8}
\\
S^1_{1} (i) &=& a^\dagger_{1} (i)  a^{\phantom\dagger}_{1} (i) , 
\label{eq9}
\\
S^{\bar{1}}_{\bar{1}} (i) &=& a^\dagger_{\bar{1}} (i)  
                              a^{\phantom\dagger}_{\bar{1}} (i) ,
\label{eq10}                              
\end{eqnarray}
where $a^\dagger_{1} (i), a^\dagger_{\bar{1}} (i)$ create magnetic excitation 
from $|0\rangle_i$ to $|1\rangle_i, |{-1}\rangle_i$, respectively. 
Here we have introduced two flavors of the boson operators. This is very
different from the usual Holstein-Primakoff transformation where
only one boson is introduced to describe the quantum fluctuation
of the magnetic order. The underlying reason is due to the particular
form of the Hamiltonian and the quantum paramagnetic ground state
that allow the excitations of the $|1\rangle_i, |{-1}\rangle_i$ 
states to be equally important. As a consequence, the excitation 
spectra for this quantum paramagnet should have eight bands,
rather than the four bands in the usual Holstein-Primakoff 
spin wave theory. Moreover, since the model has no continuous
symmetry, the magnetic excitation should be fully gapped. 

\subsection{Linear flavor wave theory}

To carry out the actual calculation of the excitation spectra, we   
replace the physical spin operators using the flavor wave transformation
and keep the Hamiltonian to the quadratic orders in the boson operators.
The resulting flavor wave Hamiltonian is given as
\begin{eqnarray}
H_{\text{fw}} = \sum_{\boldsymbol k} 
\Psi_{{\boldsymbol k}}^\dagger M({\boldsymbol k}) 
\Psi_{{\boldsymbol k}}^{\phantom\dagger} ,
\label{eq12}
\end{eqnarray}
where 
\begin{eqnarray}
\Psi_{{\boldsymbol k}}^{} &\equiv& \big( 
a_{{\boldsymbol k}01}^{\phantom\dagger} ,
a_{{\boldsymbol k}0\bar{1}}^{\phantom\dagger} ,
a_{{\boldsymbol k}11}^{\phantom\dagger} ,
a_{{\boldsymbol k}1\bar{1}}^{\phantom\dagger} ,
a_{{\boldsymbol k}21}^{\phantom\dagger} ,
a_{{\boldsymbol k}2\bar{1}}^{\phantom\dagger} ,
a_{{\boldsymbol k}31}^{\phantom\dagger} ,
a_{{\boldsymbol k}3\bar{1}}^{\phantom\dagger}, 
\nonumber \\
&& a_{\bar{\boldsymbol k}01}^{\dagger} ,
a_{\bar{\boldsymbol k}0\bar{1}}^{\dagger} ,
a_{\bar{\boldsymbol k}11}^{\dagger} ,
a_{\bar{\boldsymbol k}1\bar{1}}^{\dagger} ,
a_{\bar{\boldsymbol k}21}^{\dagger} ,
a_{\bar{\boldsymbol k}2\bar{1}}^{\dagger} ,
a_{\bar{\boldsymbol k}31}^{\dagger} ,
a_{\bar{\boldsymbol k}3\bar{1}}^{\dagger}
\big)^{T},
\nonumber \\
\end{eqnarray}
and $M({\boldsymbol k})$ is a $16\times 16$ matrix.
Here $\bar{\boldsymbol k} \equiv - {\boldsymbol k}$. 
Due to the choice of notation, $M({\boldsymbol k})$
can be written in block form as

\begin{equation}
M(\boldsymbol{k})=
\begin{pmatrix}
M_1(\boldsymbol{k}) & M_2(\boldsymbol{k})
\\
M_2^*(\boldsymbol{k}) & M_1^*(\boldsymbol{k})
\end{pmatrix},
\end{equation}
where $M_1(\boldsymbol{k})$ and $M_2(\boldsymbol{k})$ are $8\times8$ matrices 
and satisfy $M_1^\dagger(\boldsymbol{k})=M_1(\boldsymbol{k})$, $M_2^T(\boldsymbol{k})=M_2(\boldsymbol{k})$. The detailed matrix elements are listed in the Appendix~\ref{app3}. 

\begin{figure}[t]
\includegraphics[width=6cm]{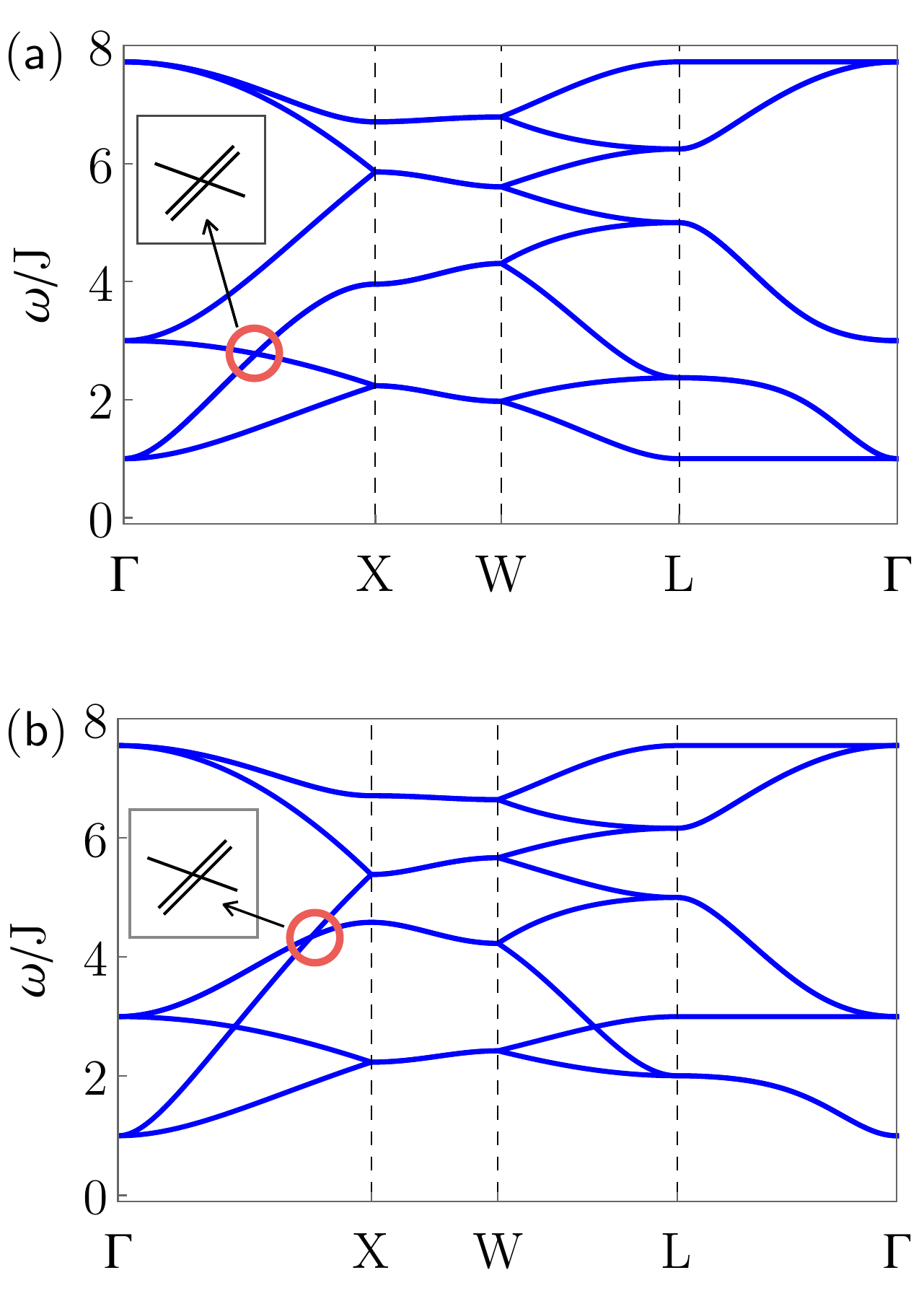}
\caption{The (gapped) magnetic excitations in the quantum
paramagnetic phase from the linear flavor wave theory.
Notice the existence of the triply degenerate nodes 
(red circle) in the spectrum, see the main text for detailed discussion.
In the inset of (a), the two-fold degenerate bands are split artificially 
for demonstration. The parameters are (a) ${D=-0.14J,D_z=5J}$; (b) ${D=0.14J,D_z=5J}$. 
The high symmetry momenta in the Brillouin zone are defined as ${\Gamma=(0, 0, 0)}$, 
${\text{X}=(0, 2\pi, 0)}$, ${\text{W}=(\pi,2\pi,0)}$, ${\text{L}=(\pi, \pi, \pi)}$.}
\label{fig3}
\end{figure}

In Fig.~\ref{fig3}, we plot the linear flavor wave dispersion for the
specific choices of the couplings within the quantum paramagnetic phase. 
As we expect, there are eight bands of the magnetic excitations that 
are fully gapped. Besides the doubled number of the bands, we notice
other unusual properties of the excitations. We find that, 
in the ${D<0}$ region of the quantum paramagnetic phase, the minima 
of the magnetic excitations develop a line of degeneracies 
from $\Gamma$ to $L$ in the momentum space. In the ${D>0}$ region of the 
quantum paramagnetic phase, the band minima of the two lowest bands touch 
at the $\Gamma$ point with an accidental two-fold degeneracy in the spin 
space. Both the momentum space degeneracy and the spin space degeneracy 
are not protected by any symmetry of the spin Hamiltonian. We expect 
the emergent degeneracy to be lifted when we go beyond the linear flavor 
wave theory and include the interaction between the flavor bosons.

\subsection{Critical properties from flavor wave theory}

\begin{figure}[b]
	\includegraphics[width=6cm]{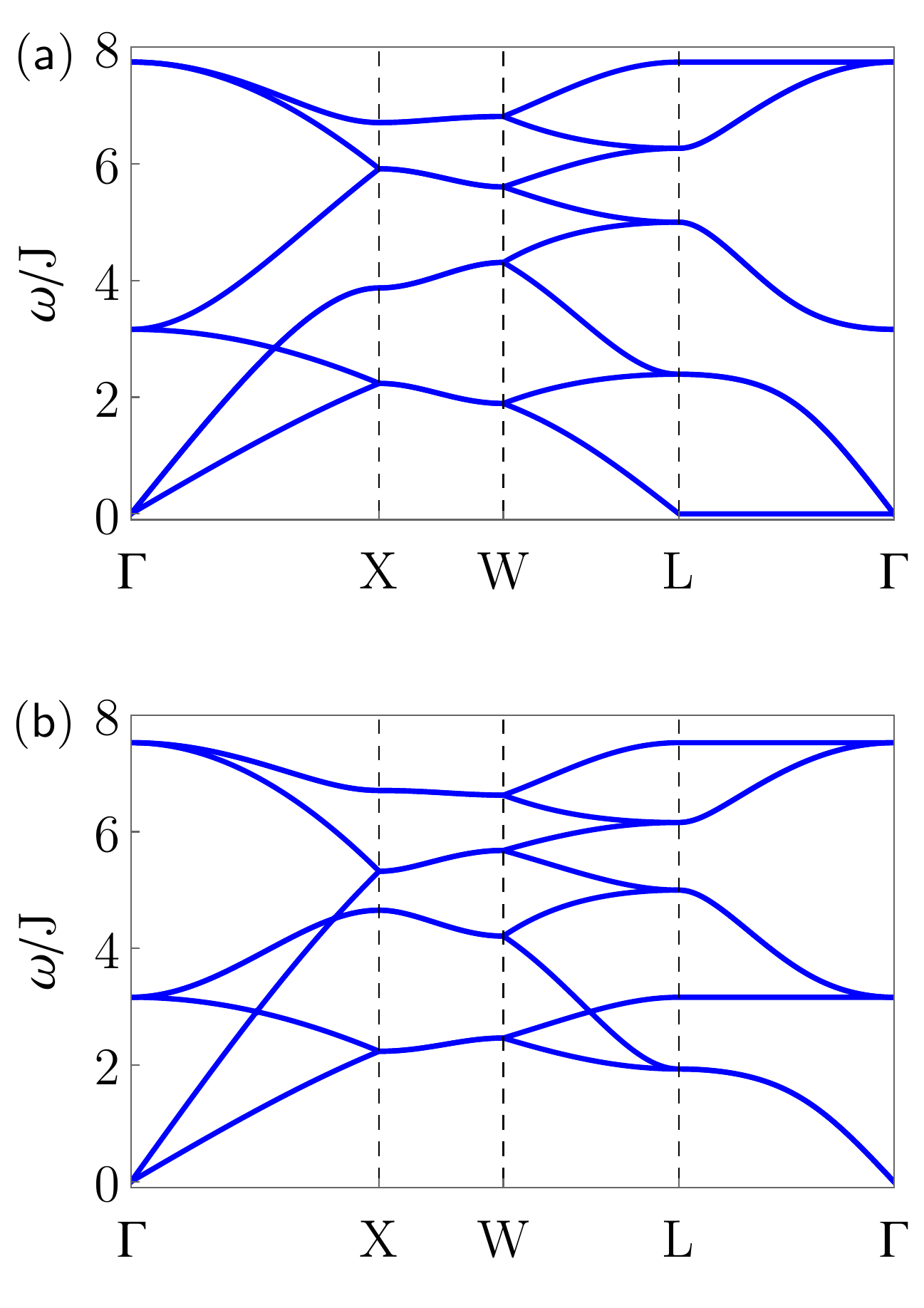}
	\caption{The magnetic excitations on the phase boundary of the quantum paramagnet, 
	obtained from the linear flavor wave theory. The excitation gap is closed. 
	The parameters are (a) ${D=-0.17J}$, ${D_z=5J}$; (b) ${D=0.17J}$, ${D_z=5J}$.}
	\label{fig4}
\end{figure}

As we further increase the exchange interaction from the quantum paramagnet, 
the gap of the magnetic excitations gradually diminishes. Eventually,  
as the gap is closed, phase transition happens and the system develops 
magnetic orders. To understand the critical properties, we examine
the transition from the flavor wave theory. In the ${D<0}$ region, 
the degenerate modes along the momentum line from $\Gamma$ to $L$ 
become critical at the same time as the gap is closed, see Fig.~\ref{fig4}(a).
Because of the line degeneracy, there is an enhanced density of states
at low energies at the criticality, and we would expect the specific heat 
${C_v \sim T^2}$ behavior at low temperatures from the mean-field theory.  
The zero-temperature limit of the specific heat should be modified 
because the fluctuations break the momentum space degeneracy 
and lead to discrete degeneracy. 
In the ${D>0}$ region, as the system approaches the criticality, only 
the $\Gamma$ point becomes critical, see Fig.~\ref{fig4}(b), and we expect 
a simple ${C_v \sim T^3}$ at the mean-field level and a logarithmic 
correction when the fluctuations beyond the mean-field are included.

\begin{figure*}[t]
	\includegraphics[width=16.8cm]{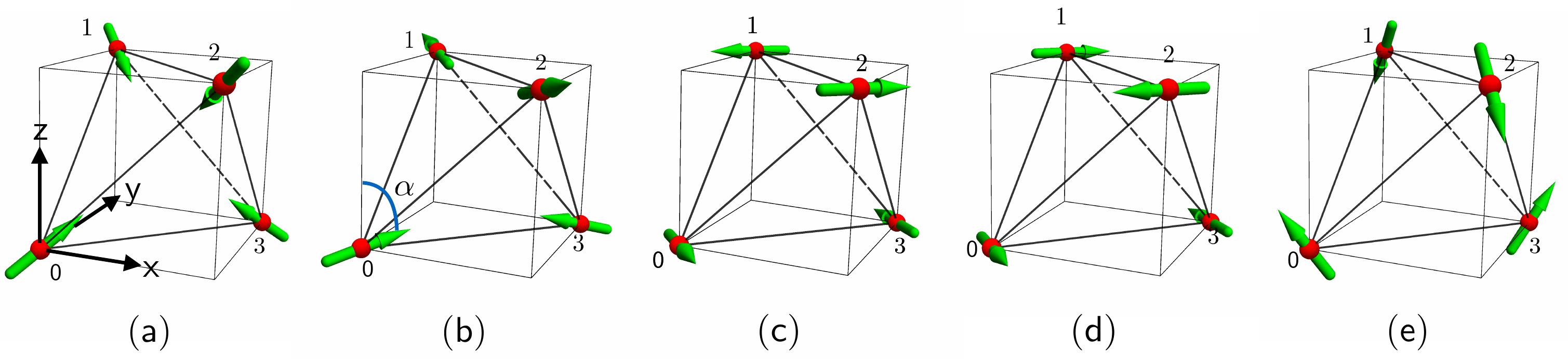}
	\caption{Representative configurations of the magnetic ordered states. (a) All-in-all-out. (b) Splayed FM. The splay angle is labeled by $\alpha$. For this configuration, $\alpha=74.2^\circ$. 
		(c) Coplanar XY AFM$_1$. (d) Coplanar XY AFM$_2$. (e) Non-coplanar XY AFM.}
	\label{fig5}
\end{figure*}

\subsection{Flavor wave excitations}

In the flavor wave excitation spectrum, there exist triply degenerate 
nodes along $\Gamma$-X and symmetry equivalent momentum directions, 
indicated by red circles in Fig.~\ref{fig3}. 

In the insets of Fig.~\ref{fig3}, we sketch that there are two-fold 
degenerate bands near the triply degenerate nodes. This two-fold band 
degeneracy is protected by a glide symmetry, which can be realized by 
a reflection in (100) plane followed by a fractional translation 
$\left({1}/{2}, {1}/{4}, {3}/{4}\right)$ in our origin choice 
(see Fig.~\ref{fig2}(a)).
This symmetry operation keeps the $\Gamma$-X line invariant and permutes 
the sublattices as $0\leftrightarrow1$ and $2\leftrightarrow3$.
Since a generic field removes the glide symmetry and lifts the two-fold 
band degeneracy, one can apply an external magnetic field to open 
a gap in the position of a triply degenerate node.

The triply degenerate nodes have been previously discussed in 
the electronic systems\cite{Bradlynaaf5037,PhysRevB.93.241202,PhysRevX.6.031003,nature22390}. 
Unlike the cases for the electronic systems where the modes at 
the nodes become unconventional quasiparticles if the Fermi 
level is tuned to the nodes, these excitations 
occur at the finite energies for the bosonic flavor waves.

We mention that in Fig.~\ref{fig3}(b), there exist doubly 
degenerate touchings along $\Gamma$-X, W-L and symmetry 
equivalent momentum directions. These touchings belong 
to a nodal surface rather than being isolated nodes, 
we will discuss their properties in future works.

\section{Mean-field theory}
\label{sec4}

To study the proximate magnetic order out of the quantum paramagnetic 
phase, one natural approach would simply follow the flavor wave theory
that we have introduced in the previous section and study the condensation 
of the critical flavor wave modes. This is certainly feasible and 
requires including the interactions between the flavor wave modes  
that lift the degeneracy of the low-energy modes. We, however,    
implement a mean-field theory in this section. This is justified  
since the system develops magnetic orders in the parameter regimes 
that we are interested. This mean-field approach works best deep 
on the ordered side. In the mean-field theory, we simply replace 
the spin operator with the mean-field order parameter
and optimize the mean-field Hamiltonian, 
\begin{eqnarray}
\langle H \rangle  
&=& \sum_{\langle ij \rangle} 
J \, {\boldsymbol m}_i \cdot {\boldsymbol m}_j + {\boldsymbol D}_{ij} 
\cdot ({\boldsymbol m}_i \times {\boldsymbol m}_j )
\nonumber \\
&& + \sum_i D_z ({\boldsymbol m}_i \cdot \hat{z}_i) ^2 , 
\end{eqnarray}
under the local constraint ${|{\boldsymbol m}_i|^2 = S^2}$. The mean-field 
ground state can then be found using the simple Luttinger-Tisza method. 
Our results are summarized and displayed in Fig.~\ref{fig1}
and Fig.~\ref{fig5}. All of these orders support an ordering wavevector 
${{\boldsymbol Q} ={\boldsymbol 0}}$ where the magnetic unit cell 
coincides with the crystal unit cell. In the following, we describe 
the magnetic orders in details. 
Since we are interested in magnetic orders in this section, our results 
will be presented from bottom to top and from left to right 
in the phase diagram of Fig.~\ref{fig1}.

\subsection{All-in all-out AFM}

In the lower left region of the phase diagram, the ``all-in all-out'' 
magnetic order is stabilized. This is understood as follows. The  
easy-axis anisotropy favors the spins to be aligned with the local
$\hat{z}$ direction, and the Heisenberg interaction requires the 
vector addition of the spins from the four sublattices to be zero. 
The Dzyaloshinskii-Moriya interaction is less obvious, but naturally
favors non-collinear spin configurations. Simple diagonalization
of the Dzyaloshinskii-Moriya interaction term directly gives 
the ``all-in all-out'' spin configuration. Therefore, all three 
interactions in the Hamiltonian are optimized by the ``all-in all-out'' 
spin configuration. Since the Dzyaloshinskii-Moriya interaction favors 
this ground state, this ``all-in all-out'' state extends further into 
the easy-plane anisotropic regime with $D_z >0$. As the local $\hat{z}$ 
direction is a three-fold rotational axis, this symmetry operation does 
not generate new ground states, and the ground state spin configuration
merely has a $\mathbb{Z}_2$ degeneracy from the time-reversal transformation.

\subsection{Splayed FM}

In the lower right region of the phase diagram, the ``splayed ferromagnet'' 
(``splayed FM'') is stabilized. One such spin configuration is given 
in Fig.~\ref{fig5}(b) and parameterized as
\begin{eqnarray}
\left\{ \begin{array}{l}
{\boldsymbol m}_0  =  (\frac{\sin\alpha}{\sqrt{2}}, \frac{\sin\alpha}{\sqrt{2}}, \cos\alpha) ,
\\[.4cm]
{\boldsymbol m}_1  =  (-\frac{\sin\alpha}{\sqrt{2}}, \frac{\sin\alpha}{\sqrt{2}}, \cos\alpha) ,
\\[.4cm]
{\boldsymbol m}_2  =  (\frac{\sin\alpha}{\sqrt{2}}, -\frac{\sin\alpha}{\sqrt{2}}, \cos\alpha) ,
\\[.4cm]
{\boldsymbol m}_3  = (-\frac{\sin\alpha}{\sqrt{2}}, -\frac{\sin\alpha}{\sqrt{2}}, \cos\alpha) ,
\end{array}
\right.
\end{eqnarray}
where ${\boldsymbol m}_{\mu}$ refers to the magnetic order on the $\mu$-th sublattice,
and the ``splay angle'' $\alpha$ is found to be
\begin{equation}
\alpha =\arctan \frac{ D_z' - [8D_z^2 + {D'_z}^2]^{ \frac{1}{2} }}{2\sqrt{2}D_z},
\end{equation}
here ${D'_z \equiv D_z -12 J -3\sqrt{2} D}$. There is a ferromagnetic component $\cos\alpha$
along the global $z$ direction.

Other equivalent ground state spin configurations can be obtained 
by lattice symmetry operations, and we have the other ground states as 
\begin{eqnarray}
\left\{ \begin{array}{l}
{\boldsymbol m}_0  = (\frac{\sin\alpha}{\sqrt{2}}, \cos\alpha, \frac{\sin\alpha}{\sqrt{2}}) ,
\\[.4cm]
{\boldsymbol m}_1  = (-\frac{\sin\alpha}{\sqrt{2}}, \cos\alpha, \frac{\sin\alpha}{\sqrt{2}}) ,
\\[.4cm]
{\boldsymbol m}_2  =  (-\frac{\sin\alpha}{\sqrt{2}}, \cos\alpha, -\frac{\sin\alpha}{\sqrt{2}}) ,
\\[.4cm]
{\boldsymbol m}_3  = (\frac{\sin\alpha}{\sqrt{2}}, \cos\alpha, -\frac{\sin\alpha}{\sqrt{2}}) ,
\end{array}
\right.
\end{eqnarray}
and 
\begin{eqnarray}
\left\{ \begin{array}{l}
{\boldsymbol m}_0  = (\cos\alpha, \frac{\sin\alpha}{\sqrt{2}}, \frac{\sin\alpha}{\sqrt{2}}) ,
\\[.4cm]
{\boldsymbol m}_1  = (\cos\alpha, -\frac{\sin\alpha}{\sqrt{2}}, -\frac{\sin\alpha}{\sqrt{2}}) ,
\\[.4cm]
{\boldsymbol m}_2  = (\cos\alpha, -\frac{\sin\alpha}{\sqrt{2}}, \frac{\sin\alpha}{\sqrt{2}}) ,
\\[.4cm]
{\boldsymbol m}_3  =  (\cos\alpha, \frac{\sin\alpha}{\sqrt{2}}, -\frac{\sin\alpha}{\sqrt{2}}).
\end{array}
\right.
\end{eqnarray}

Together with the time reversal symmetry, there exist a 
${\mathbb{Z}_3 \times {\mathbb Z}_2}$ degeneracy. 
This state supports a weak 
ferromagnetism along one cubic axis and antiferromagnetism 
in the remaining two directions. Clearly, when $|D_z|$ is dominant,
the spins should be aligned with the local $\hat{z}$ direction,
and the Dzyaloshinskii-Moriya interaction then favors ``two-in two-out''
spin configurations in this case. 

In the strong $D_z$ limit, the splay angle ${\alpha\approx54.7^\circ}$, 
and the ground state is exactly the ``two-in two-out'' spin ice
configurations. In contrast, in the weak $D_z$ limit, ${\alpha=90^\circ}$ 
and the ground state becomes coplanar.
This means the ``two-in two-out'' spin ice
configurations are smoothly connected to coplanar states 
in this ``splayed FM'' regime.

In general, in this parameter regime, the interactions cannot 
be optimized simultaneously. However, taking three interactions 
together, we are able to find the ``splayed FM'' as the ground state. 
This ``splayed FM'' was actually proposed for the well-known quantum spin ice 
candidate materials Yb$_2$Sn$_2$O$_7$ and Yb$_2$Ti$_2$O$_7$~\cite{PhysRevLett.110.127207,PhysRevLett.119.057203}, 
so we adopt the name from there. We note that the splay angle $\alpha$ can 
only take value from $54.7^\circ$ to $90^\circ$ for the ``splayed FM'' 
regime with antiferromagnetic Heisenberg exchange. 
When the Heisenberg exchange becomes ferromagnetic, $\alpha$ 
can take a larger parameter regime (see Appendix~\ref{app5}).

\subsection{Coplanar XY AFM$_1$}
\label{ssec3}

In the upper left region of the phase diagram, we obtain a coplanar     
antiferromagnetic spin ground state and dub it ``coplanar XY AFM$_1$''. 
Here `XY' refers to the $xy$ plane of the local coordinate system. 
One such spin state is depicted in Fig.~\ref{fig5}(c) and is given as 
\begin{eqnarray}
\left\{ \begin{array}{l}
{\boldsymbol m}_0  =  \frac{1}{\sqrt{2}} (1, \bar{1}, 0) ,
\\[.4cm]
{\boldsymbol m}_1  =  \frac{1}{\sqrt{2}} (\bar{1}, \bar{1}, 0) ,
\\[.4cm]
{\boldsymbol m}_2  =  \frac{1}{\sqrt{2}} (1,{1},{0}) ,
\\[.4cm]
{\boldsymbol m}_3  =  \frac{1}{\sqrt{2}} (\bar{1}, {1},0) .
\end{array}
\right.
\label{eq20}
\end{eqnarray}
The spins are perpendicular to the local $\hat{z}$ direction of the 
relevant sublattice and orient antiferromagnetically within the same 
plane globally. This explains the use of the ``coplanar XY AFM$_1$''. 
This ``coplanar XY AFM$_1$'' ground state occurs when ${D_z > \sqrt{2} |D|}$
as one further increases the easy-plane anisotropy from the ``all-in all-out''
phase. This ``coplanar XY AFM$_1$'' phase is in the easy-plane anisotropic 
limit, and the spins prefer to orient in the local $xy$ plane. The in-plane 
spin configuration is able to content both the easy-plane spin anisotropy
and the Heisenberg exchange. Since it is known from the previous subsection 
that the Dzyaloshinskii-Moriya interaction is optimized by the 
``all-in all-out'' state for ${D<0}$. The particular spin configuration of 
the ``coplanar XY AFM$_1$'' state is obtained because the easy-plane anisotropy 
wins over the Dzyaloshinskii-Moriya interaction such that the Dzyaloshinskii-Moriya 
interaction is optimized within the manifold of coplanar spin configurations only.

Applying the lattice symmetry operations, we generate two equivalent spin
configurations with 
\begin{eqnarray}
\left\{ \begin{array}{l}
{\boldsymbol m}_0  =  \frac{1}{\sqrt{2}} (0, 1, \bar{1}) ,
\\[.4cm]
{\boldsymbol m}_1  =  \frac{1}{\sqrt{2}} (0, \bar{1}, {1}) ,
\\[.4cm]
{\boldsymbol m}_2  =  \frac{1}{\sqrt{2}} (0, \bar{1},\bar{1}) ,
\\[.4cm]
{\boldsymbol m}_3  =  \frac{1}{\sqrt{2}} (0, {1}, {1}) ,
\end{array}
\right.
\end{eqnarray}
and 
\begin{eqnarray}
\left\{ \begin{array}{l}
{\boldsymbol m}_0  =  \frac{1}{\sqrt{2}} (1, 0, \bar{1}) ,
\\[.4cm]
{\boldsymbol m}_1  =  \frac{1}{\sqrt{2}} (\bar{1},0, \bar{1}) ,
\\[.4cm]
{\boldsymbol m}_2  =  \frac{1}{\sqrt{2}} (\bar{1}, 0, {1}) ,
\\[.4cm]
{\boldsymbol m}_3  =  \frac{1}{\sqrt{2}} ({1}, 0,  {1}) .
\end{array}
\right.
\end{eqnarray}
Again from the time reversal symmetry, we have a $\mathbb{Z}_3 \times \mathbb{Z}_2$
degeneracy for the ground state.

\subsection{Coplanar XY AFM$_2$}
\label{ssec4}

\begin{figure}[t]
	\includegraphics[width=8.5cm]{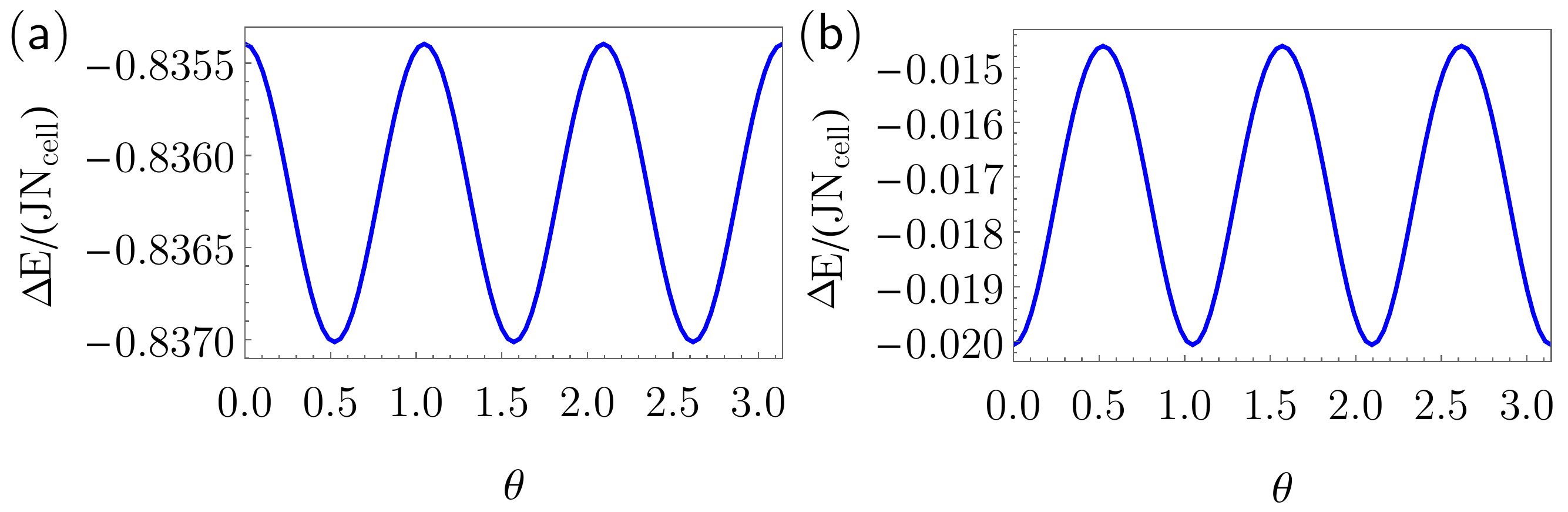}
	\caption{Order by quantum disorder in the upper right region of the phase diagram. 
		Coplanar XY AFM$_2$ and Non-coplanar XY AFM are separated by different 
		(a) In coplanar XY AFM$_2$, the minima of zero point energy are realized 
		at $\theta = {n\pi}/{3}+{\pi}/{6}$ for ${n\in \mathbb{Z}}$. 
		(b) In non-coplanar XY AFM,  the minima of zero point energy are realized at
		$\theta = {n\pi}/{3}$ for ${n\in \mathbb{Z}}$. 
		The parameters are (a) ${D=0.1J,D_z=0.7J}$; (b)${D=J,D_z=0.5J}$.}
	\label{fig6}
\end{figure}

In the upper right region (both the ``coplanar XY AFM$_2$'' and 
``non-coplanar XY AFM'') of the phase diagram, we find an extensively degenerate 
mean-field ground state, and all the three interactions are optimized
at the same time. The extensive degeneracy is parametrized by a $U(1)$ 
angular variable $\theta$, and the ground state spin configuration is given as 
\begin{eqnarray}
{\boldsymbol m}_{\mu} = \hat{x}_{\mu} \cos \theta + \hat{y}_{\mu} \sin \theta ,
\end{eqnarray}
with ${\theta \in [0,2\pi)}$. 
Our spin Hamiltonian does not have any continuous symmetry, thus the 
continuous degeneracy is not the symmetry property of the Hamiltonian
but is accidental. We expect this continuous degeneracy to be lifted 
by quantum fluctuation. This quantum order by disorder effect
has been previously explored in the effective spin-1/2 pyrochlore 
material Er$_2$Ti$_2$O$_7$~\cite{PhysRevLett.109.167201,PhysRevLett.109.077204,PhysRevB.89.140403}. 
We here study this quantum mechanical effect in the spin-1 pyrochlore system. 
We first introduce the Holstein-Primakoff transformation for the 
spin operators,
\begin{eqnarray}
&& {\boldsymbol S}_i \cdot {\boldsymbol m}_i =
S-b^\dagger_i b^{\phantom\dagger}_i ,
\\
&& {\boldsymbol S}_i \cdot \hat{z}_i = 
\frac{\sqrt{2S}}{2} (b^{\phantom\dagger}_i + b^\dagger_i) ,
\\
&& {\boldsymbol S}_i \cdot ({\boldsymbol m}_i\times\hat{z}_i) =
\frac{\sqrt{2S}}{2i} (b^{\phantom\dagger}_i - b^\dagger_i).
\end{eqnarray}
Substituting the spin operators with the Holstein-Primakoff bosons 
and keeping the boson terms up to quadratic order, 
we have the linear spin wave Hamiltonian (see Appendix~\ref{app4}), 
\begin{eqnarray}
H_{\text{sw}} &=& \sum_{\boldsymbol k} \sum_{\mu\nu} 
\bigg[ \frac{D_z}{2}\delta_{\mu\nu} + A_{\mu\nu}({\boldsymbol k})
b_{{\boldsymbol k}\mu}^\dagger b_{{\boldsymbol k}\nu}^{\phantom\dagger} 
\nonumber \\
&& +\left(B_{\mu\nu}({\boldsymbol k}) 
b_{{\boldsymbol k}\mu}^\dagger b_{-{\boldsymbol k}\nu}^\dagger 
+ \text{h.c.}\right) \bigg] + E_{\text{mf}} ,
\label{eq27}
\end{eqnarray}
where $E_{\text{mf}}$ is the mean-field energy 
of the ground state. The quantum zero point energy is found to be 
\begin{eqnarray}
\Delta E = \sum_{\boldsymbol k} \sum_{\mu} \frac{1}{2} 
\Big[\omega_\mu({\boldsymbol k})-A_{\mu\mu}({\boldsymbol k})+D_z\Big] ,
\end{eqnarray}
here $\omega_{\mu} ({\boldsymbol k})$ is the spin wave excitation.
In Fig.~\ref{fig6}, we plot the quantum zero point energy and find that 
the minima are realized at
\begin{equation}
\theta = \frac{n\pi}{3}+\frac{\pi}{6} ,
\end{equation}
for ${n\in \mathbb{Z}}$, see Fig.~\ref{fig6}(a). One such spin configuration 
is displayed in Fig.~\ref{fig5}(d), and all the spins orient antiferromagnetically 
within the same plane.  We dub this phase ``coplanar XY AFM$_2$''.

\subsection{Non-coplanar XY AFM} 
\label{ssec5}

In the remaining part of the upper right region in the phase diagram,
quantum fluctuation leads to different ground state spin configurations.
As we plot in Fig.~\ref{fig6}(b), the minima of the zero-point energy are realized at 
\begin{equation}
\theta = \frac{n\pi}{3}
\end{equation}
for ${n\in \mathbb{Z}}$. One such spin configuration is displayed 
in Fig.~\ref{fig5}(e), and all the spins orient antiferromagnetically 
but are not in the same plane. This phase is dubbed ``non-coplanar XY AFM''.

\subsection{Phase boundaries between ordered phases}

Here we explain the phase boundaries between different ordered phases. 
The phase boundary between ``coplanar XY AFM$_2$'' and ``non-coplanar XY AFM'' is
numerically determined by finding the minima of the quantum zero-point energy.
The other phase boundaries are determined by energy competition between different 
interactions at the mean-field level and understood from the connection to 
the Heisenberg point. Since the order parameter is disconnected between 
different ordered phases, all the phase transitions across the boundaries 
are expected to be first order.

We start from the phase boundary between ``all-in all-out'' and ``splayed FM''. This
boundary is defined by the curve
\begin{equation}
|D_z|=\frac{9D(D-\sqrt{2}J)}{2\sqrt{2}D-J}.
\end{equation}

``All-in all-out'' and ``coplanar XY AFM$_1$'' are separated by 
the line ${D_z=\sqrt{2}|D|}$. The remaining two boundaries are 
the line ${D=0, D_z>0}$, separating ``coplanar XY AFM$_1$'' from
``coplanar XY AFM$_2$'' and ``non-coplanar XY AFM'', and the line 
${D_z=0, D>0}$, separating ``coplanar XY AFM$_2$'' from ``splayed FM''. 
There is enlarged mean-field ground state manifold on these three lines. 
If the spin configurations of two neighboring phases, say 
$\boldsymbol{m}_{i}^1$ and $\boldsymbol{m}_{i}^2$ respectively, 
are orthogonal with $\boldsymbol{m}_{i}^1\cdot\boldsymbol{m}_{i}^2=0$ 
for each sublattice, one can readily construct a ground state manifold
with $U(1)$ degeneracy on the phase boundary, written as
\begin{equation}
\boldsymbol{m}_{i}=\cos\varphi~ \boldsymbol{m}_{i}^1 
		+ \sin\varphi~ \boldsymbol{m}_{i}^2,
\end{equation}
where ${\varphi\in [0,2\pi)}$ is an angular variable.
In the Appendix~\ref{app6}, we discuss the ground state
and the order by quantum disorder effect on these phase boundaries.

\subsection{Phase boundaries to the quantum paramagnet}

As we have explained in the beginning of this section, there are two approaches 
to establish the magnetic orders of this system. One approach is to start from 
the quantum paramagnet by condensing the flavor wave boson. The other approach 
is to implement the mean-field theory and is adopted in this section. 
To build the connection between the proximate magnetic orders with the 
quantum paramagnet within the latter approach, one could apply the Weiss
type of mean-field theory by assuming the proximate magnetic order as the 
mean-field ansatz and examine the disappearance of the magnetic orders. 
This treatment necessarily finds a direct transition between 
the proximate magnetic order and the quantum paramagnet, and does not 
provide more qualitatively new information than the former approach. 
The current phase boundary is established from the former approach. 
Intermediate phases such as the chiral liquid phase 
with a finite vector chirality order may be stabilized by the 
flavor wave interaction that is not considered in this work.  

For the current phase diagram, we explain the connection between the 
proximate orders and the quantum paramagnet. On the upper left part 
of the phase diagram, as we show in previous section, the flavor wave 
excitation has a line degeneracy in the momentum space from $\Gamma$ 
to $L$. This momentum space degeneracy is accidental and is also found 
in the mean-field treatment if one penalizes the local constraint for 
the magnetic orders. Since the candidate magnetic states with the 
wavevectors other than the $\Gamma$ point cannot satisfy the local 
constraint, thus only the coplanar state that is discussed 
in Sec.~\ref{ssec3} survives. On the upper right part of the phase diagram,
the band minimum of the flavor wave excitation in the quantum paramagnet
appears at the $\Gamma$ point and has two degenerate modes. 
The degenerate modes, when they are condensed, lead to the continuous
$U(1)$ degeneracy within the manifold of these two modes at the mean-field 
description. This $U(1)$ degeneracy is precisely the $U(1)$ degeneracy
that is discussed in Sec.~\ref{ssec4} and Sec.~\ref{ssec5}.

\subsection{Topological magnons and spin wave excitations of the ordered phases}

\begin{figure}[t]
	\includegraphics[width=8.4cm]{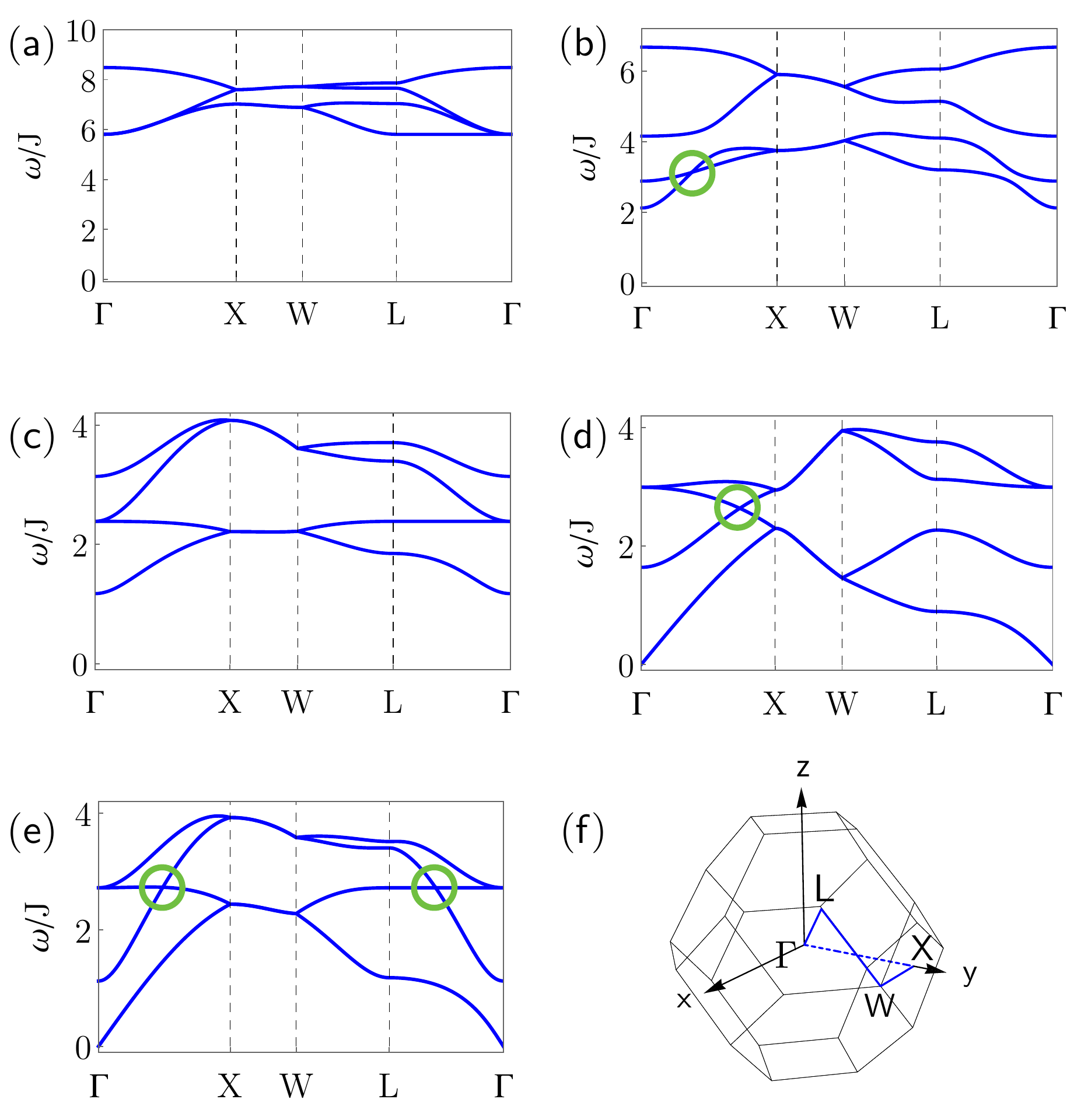}
	\caption{Spin wave excitations of the ordered phases.
		The parameters are (a) ${D=-J,D_z=0}$ (all-in all-out); 
		(b) ${D=J,D_z=-0.3J}$ (splayed FM); 
		(c) ${D=-0.3J,D_z=0.6J}$ (coplanar XY AFM$_1$);
		(d) ${D=0.1J,D_z=J}$ (non-coplanar XY AFM);
		(e) ${D=0.5J,D_z=0.1J}$ (coplanar XY AFM$_2$).
		In (f) we plot the Brillouin zone of the pyrochlore lattice and 
		label the high symmetry points.
	    }
	\label{fig11}
\end{figure}

In Fig.~\ref{fig11}, we plot the spin wave excitation of each 
ordered phase along high symmetry lines in Brillouin zone. 
As expected, the spectra in Fig.~\ref{fig11}(a)(b)(c) are 
fully gapped while in Fig.~\ref{fig11}(d)(e),
there are gapless pseudo-Goldstone modes at $\Gamma$, 
reflecting the continuous $U(1)$ degeneracy in the
mean-field ground state manifold. Since the degeneracy 
is accidental, a small gap is expected when we go 
beyond the linear spin wave approximation.

\begin{figure}[h]
\includegraphics[width=0.49\textwidth]{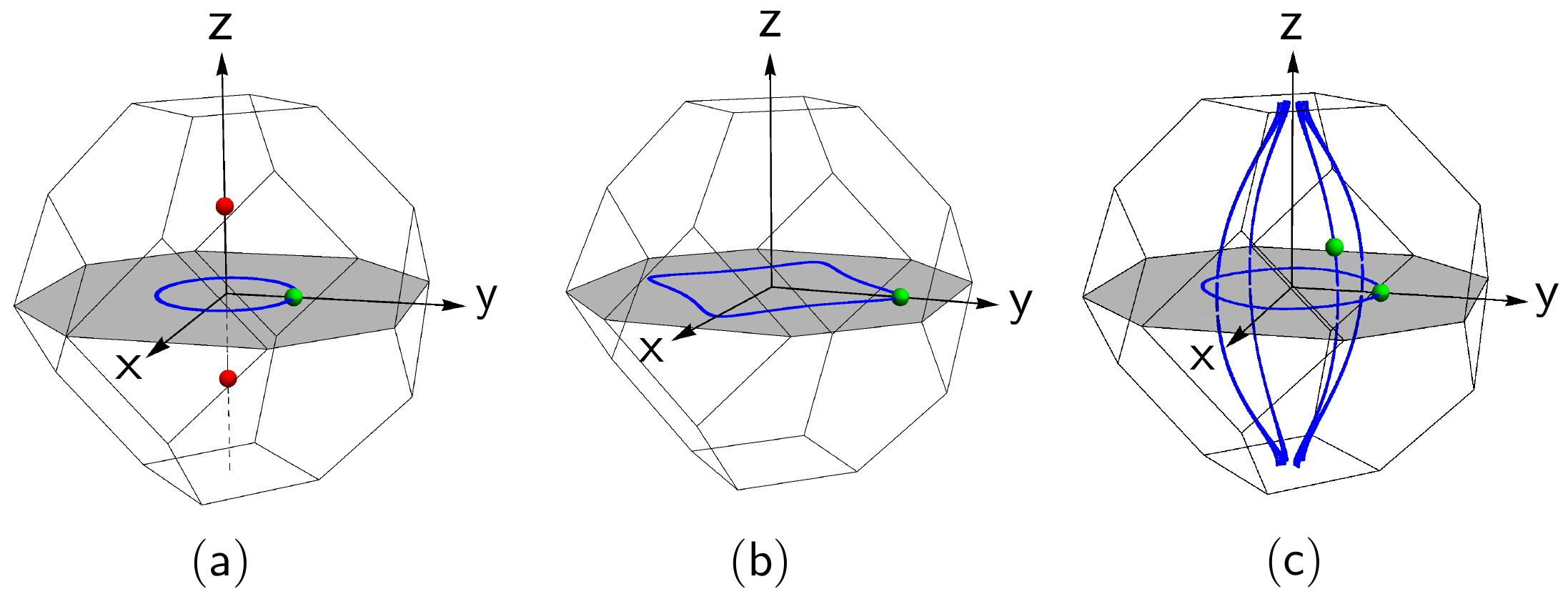}
\caption{The nodal lines and Weyl nodes of the spin wave excitation. 
(a) For the same parameters as in Fig.~\ref{fig11}(b), there is a nodal 
contour in the (001) plane (gray) of the reciprocal space. The band 
touching showed in Fig.~\ref{fig11}(b) is labeled by a green dot here. 
Moreover, there exists a pair of isolated Weyl nodes along $z$ axis, 
labeled by red dots.
(b) For the same parameters as in Fig.~\ref{fig11}(d), there is a nodal 
line in the (001) plane (gray) of the reciprocal space too. Again the 
band touching showed in Fig.~\ref{fig11}(d) is labeled by a green dot.
(c) For the same parameters as in Fig.~\ref{fig11}(e), the nodal lines 
form a cage-like structure. One nodal contour is located in the (001) 
plane (gray) and intersects with the other four nodal lines, of which 
two are located in the (110) plane and the other two are located in 
the (1$\bar{1}$0) plane. The two band touchings showed in Fig.~\ref{fig11}(e) 
are labeled by green dots.
}
\label{fig14}
\end{figure}

We further explore the topological spin wave modes 
in the spectrum. Besides the Weyl nodes (see Fig.~\ref{fig14}), 
we find extra doubly degenerate band touchings, 
labeled by green circles. These touchings 
belong to certain nodal lines (see Fig.~\ref{fig14}). Since 
these magnon excitations are bosonic, they occur at the 
finite energies. 
These topological magnons~\cite{Weylmagnon,PhysRevLett.117.157204,PhysRevB.95.085132,2399-6528-1-2-025007,0295-5075-117-3-37006,PhysRevB.94.075401,FangLi} 
are magnetic analogues of the the electronic topological 
semimetals~\cite{Wan2011,PhysRevB.84.235126}. 

\section{Discussion}
\label{sec5}

\subsection{Summary of theoretical results} 

In this paper, we have proposed a generic spin model to describe  
the interacting spin-one moments on the pyrochlore lattice. We have 
established a global phase diagram with very rich phases for 
this model using several different and complementary methods.
The magnetic ordered states are understood from both the mean 
field theory and the instability of the quantum paramagnetic  
phase. The relations between different phases are further 
clarified. Both the magnetic structures of the ordered phases 
and the corresponding elementary excitations are carefully
studied. We point out the existence of degenerate and topological 
excitations. While these results are valid within the 
approximation that we made, we would like to point out the 
caveat of our theoretical results. We expect that our results 
break down when the system approaches the Heisenberg limit.  
Thus, the phases in the vicinity of the Heisenberg model of Fig.~\ref{fig1}
are expected to altered, and more quantum treatment is needed. 
The ground state for the pyrochlore lattice Heisenberg model 
is one of the hardest problems in quantum magnetism. 
The early theoretical attempts provide insights
for the classical limit~\cite{PhysRevB.58.12049,PhysRevLett.80.2929}. 
Due to the extensive classical ground state degeneracy, the quantum 
fluctuation is deemed to be very strong when the quantum nature of 
the spins is considered. Moreover, there should be fundamental distinctions 
between the spin-1/2 and the spin-1 Heisenberg models.

\begin{table*}[t]
\centering
\begin{tabular}{cccccc}
\hline\hline
materials &  magnetic ions & $\Theta_{\text{CW}}$ & magnetic transitions & magnetic structure & refs
\\
NaCaNi$_2$F$_7$ & Ni$^{2+}$($3d^8$) & $-129$K & glassy transition at 3.6K &  spin glass & \onlinecite{PhysRevB.92.014406}
\\
Y$_2$Ru$_2$O$_7$ & Ru$^{4+}$($4d^4$) & $-1250$K & AFM transition at 76K & 
canted AFM ${{\boldsymbol Q}={\boldsymbol 0}}$ & \onlinecite{PhysRevB.74.104425}
\\
Tl$_2$Ru$_2$O$_7$ & Ru$^{4+}$($4d^4$) & $-956$K & structure transition at 120K & gapped paramagnet & \onlinecite{nmat1605}
\\
Eu$_2$Ru$_2$O$_7$ & Ru$^{4+}$($4d^4$) &  - & Ru order at 118K & Ru order & \onlinecite{Eu2Ru2O7}
\\
Pr$_2$Ru$_2$O$_7$ & Ru$^{4+}$($4d^4$), Pr$^{3+}$($4f^2$) &  $ -224$K & Ru AFM order at 162K & Ru AFM order & \onlinecite{PrRuO,ZOUARI200943}
\\
Nd$_2$Ru$_2$O$_7$ & Ru$^{4+}$($4d^4$), Nd$^{3+}$($4f^3$) & $-168$K & Ru AFM order at 143K & Ru AFM order & 
\onlinecite{0953-8984-25-18-186004}
\\
Gd$_2$Ru$_2$O$_7$ & Ru$^{4+}$($4d^4$), Gd$^{3+}$($4f^7$) &  $-10$K  & Ru AFM order at 114K & Ru AFM order
${{\boldsymbol Q}={\boldsymbol 0}}$      
& \onlinecite{PhysRevB.75.064426}
\\
Tb$_2$Ru$_2$O$_7$ & Ru$^{4+}$($4d^4$), Tb$^{3+}$($4f^8$) &  $-16$K & Ru AFM order at 110K & Ru AFM order 
${{\boldsymbol Q}={\boldsymbol 0}}$
& \onlinecite{Chang2010}
\\
Dy$_2$Ru$_2$O$_7$ & Ru$^{4+}$($4d^4$), Dy$^{3+}$($4f^9$) &  $-10$K & Ru AFM order at 100K & Ru AFM order
& \onlinecite{DyRuO}
\\
Ho$_2$Ru$_2$O$_7$ & Ru$^{4+}$($4d^4$), Ho$^{3+}$($4f^{10}$) &  $-4$K & Ru AFM order at 95K & Ru FM order ${{\boldsymbol Q}={\boldsymbol 0}}$
& \onlinecite{PhysRevLett.93.076403,B110596P}
\\
Er$_2$Ru$_2$O$_7$ & Ru$^{4+}$($4d^4$), Er$^{3+}$($4f^{11}$) & $-16$K  & Ru AFM order at 92K & Ru AFM order ${{\boldsymbol Q}={\boldsymbol 0}}$
& \onlinecite{0953-8984-21-43-436004,ErRuO}
\\
Yb$_2$Ru$_2$O$_7$ & Ru$^{4+}$($4d^4$), Yb$^{3+}$($4f^{13}$) &  - & Ru AFM order at 83K & Ru AFM order 
& \onlinecite{B110596P}
\\
Y$_2$Mo$_2$O$_7$  & Mo$^{4+}$($4d^2$) & $-200$K & Mo spin glass at 22K & Mo spin glass & 
\onlinecite{PhysRevLett.87.177201,PhysRevLett.118.067201,PhysRevB.89.054433,PhysRevB.54.9019}
\\
Lu$_2$Mo$_2$O$_7$ & Mo$^{4+}$($4d^2$) & $-160$K & Mo spin glass at 16K & Mo spin glass & 
\onlinecite{PhysRevLett.113.117201}
\\
Tb$_2$Mo$_2$O$_7$ & Mo$^{4+}$($4d^2$), Tb$^{3+}$($4f^8$) & $20$K & spin glass at 25K & spin glass &
\onlinecite{0953-8984-23-16-164214,PhysRevB.81.224405,PhysRevB.78.220405}
\\
\hline\hline
\end{tabular}
\caption{A list of candidate spin-one pyrochlore materials. 
         The null entry means that the data is not available. }
         \label{table1}
\end{table*}

\subsection{Survery of spin-one pyrochlore materials} 

There have already been several spin-one pyrochlore materials in the 
literature. We start with from the Ni-based pyrochlore material 
NaCaNi$_2$F$_7$~\cite{PhysRevB.92.014406}. This material has 
a $-129$K Curie-Weiss temperature, and no features of spin orderings  
are observed in the thermodynamic measurement until a spin glassy transition 
at 3.6K. The spin glassy transition is evidenced by the bifurcation in the 
magnetic susceptibility between the zero-field-cooled and field-cooled results. 
The magnetic entropy saturates to $Rln2$ when the temperature is increased 
to $70$K~\cite{PhysRevB.92.014406}. The highest temperature $70$K in 
the entropy measurement is probably not too large to exhaust the 
actual magnetic entropy as the Curie-Weiss temperature is $-129$K. 
If one takes this entropy result, this magnetic entropy differs from the simple 
expectation for the spin-1 local moment and indicates a significant easy-axis 
spin anisotropy that reduces the total magnetic entropy. In this case, 
based on our phase diagram in Fig.~\ref{fig1}, there would be magnetic 
orders. It is possible that the exchange randomness becomes important 
at low temperatures and drives the system into a spin glassy state instead. 
Since the glassy transition occurs at very low 
temperatures, the spin physics and dynamics at higher temperatures and 
energy scales are probably less influenced by the exchange randomness. 
If the current entropy result is not reliable due to the small upper 
temperature limit, one could extend the entropy measurement further 
in the temperature to see if one can exhaust the spin-1 magnetic entropy.  
In any case, to test the relevance of the model Hamiltonian, it can be helpful
to measure the spin correlation in the momentum space with neutron scattering
and compare with the theoretical results. Since our spin model contains the 
spin space anisotropy in addition to the momentum space due to the single-ion 
anisotropy and Dzyaloshinskii-Moriya interaction, it is also quite useful to 
carry out the polarized neutron scattering measurement on the single-crystalline
sample to detect the spin correlation function in the spin space. A very recent 
neutron scattering experiment was actually implemented on the single crystal sample.
The general features of the spin correlation seem to be well captured by the first neighbor 
Heisenberg model with much weaker further neighbor interactions~\cite{Collin2017}. 

In fact, there exists a simple and useful recipe to estimate the 
Dzyaloshinskii-Moriya interaction but not the single-ion spin anisotropy. 
The effective magnetic moment of the Ni ion in NaCaNi$_2$F$_7$ is 
found to be $3.7\mu_{\text{B}}$ from the susceptibility data from 
5K to 300K~\cite{PhysRevB.92.014406}. This deviates from $2.82\mu_{\text{B}}$
for the pure ${S=1}$ moment in the atomic limit, and this deviation 
is due to the spin-orbit coupling. It is known that the deviation 
$\Delta g$ of the Land\'{e} $g$ factor is related to the Dzyaloshinskii-Moriya
interaction~\cite{PhysRev.120.91} with ${\Delta g/g  \sim |\boldsymbol{D}_{ij}|/J}$. 
This suggests that the Dzyaloshinskii-Moriya interaction may be up to 20-30\% 
of the Heisenberg exchange in NaCaNi$_2$F$_7$. This suggestion seems to be inconsistent 
with the conclusion that the system is described by the Heisenberg model in
Ref.~\onlinecite{Collin2017}. If the latter is true, there should be 
an unknown cancellation mechanism in the exchange paths that suppress 
the Dzyaloshinskii-Moriya interaction. If the Dzyaloshinskii-Moriya 
interaction is sizable, its effect would appear in the low-temperature
magnetic properties. 

\begin{figure}[b]
	\includegraphics[width=7cm]{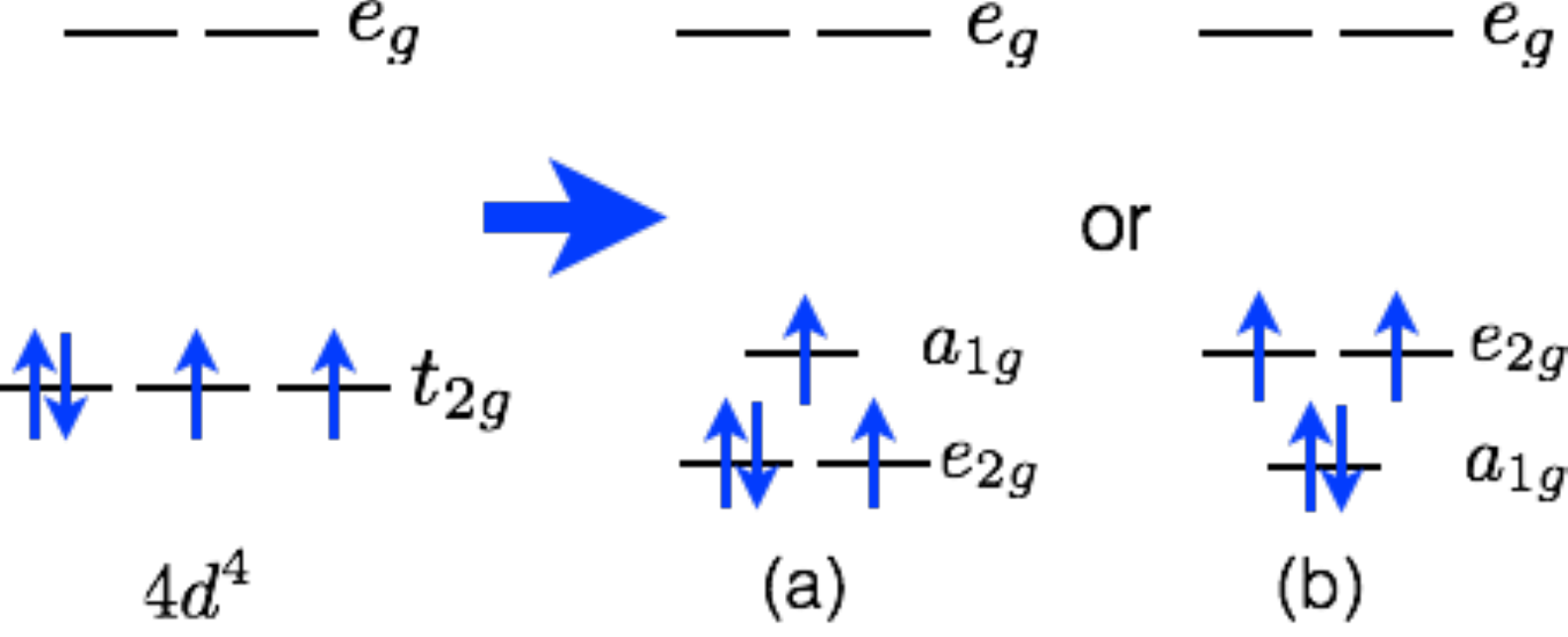}
	\caption{The orbital occupations for $4d^4$ electron configuration. 
	Under the trigonal distortion, the three-fold degenerate $t_{2g}$
	orbitals are splitted into $a_{1g}$ and two-fold degenerate $e_{2g}$ states. 
	There are two electron occupation configurations here. (a) has an unquenched
	orbital degree of freedom. }
	\label{figx1}
\end{figure}

Other existing spin-1 pyrochlore materials are the Ru-based pyrochlore 
A$_2$Ru$_2$O$_7$ and the Mo-based pyrochlore A$_2$Mo$_2$O$_7$. Both of 
them are discussed and summarized in a very nice review paper~\cite{RevModPhys.82.53} 
by Gardner, Gingras and Greedan. In both systems, the A site can be a rare-earth ion
or a non-magnetic ion with no moments. In the former case, the coupling
between the rare-earth moments and the Ru/Mo moments may be important,
and the rare-earth magnetism also contributes to the magnetic properties
of the system. If the Ru-Ru interaction is the dominant one, one may    
first consider the magnetic physics of the Ru subsystem. In the latter  
case and also for A=Eu, one only needs to consider the Ru/Mo magnetism. 

The Ru$^{4+}$ ion has a $4d^4$ electron configuration, and the electrons occupy 
the lower $t_{2g}$ orbitals. Although the atomic spin-orbit coupling is still 
active due to the partially filled $t_{2g}$ manifold, the Hund's coupling could 
suppress the effect of the spin-orbit coupling for the $4d^4$ electron configuration. 
If the spin-orbit coupling is truly dominant over the Hund's coupling, 
a quenched local moment would be obtained. Since these are $4d$ electrons, 
we expect the spin-orbit coupling could just be moderate compared to 
the Hund's coupling. From the experimental result of a spin-1 moment for 
the Ru$^{4+}$ ion, it is reasonable to take the view of a moderate spin-orbit
coupling. Moreover, 
as we show in Fig.~\ref{figx1}, there can be two different occupation
configurations after one includes the trigonal distortion. 
Fig.~\ref{figx1}a has an orbital degeneracy, while Fig.~\ref{figx1}b 
has no orbital degeneracy. The prevailing view of spin-only 
moment~\cite{RevModPhys.82.53} for the Ru$^{4+}$ ion supports 
the choice of Fig.~\ref{figx1}b. Moreover, due to different orbital 
occupation configurations and the realization of the spin-orbit coupling
for the Ru$^{4+}$ ion, although the model stays the same as Eq.~\eqref{eq1},
the single-ion anisotropy and the Dzyaloshinskii-Moriya interaction
would have different relations from the ones in Eqs.~\eqref{Eqdm} and ~\eqref{Eqion}.

As we show in Table~\ref{table1}, almost all materials in the A$_2$Ru$_2$O$_7$ family 
develop magnetic orders except Tl$_2$Ru$_2$O$_7$. We start from the materials 
with pure Ru moments. The canted AFM state, that was found 
for Y$_2$Ru$_2$O$_7$ in Ref.~\onlinecite{PhysRevB.74.104425}, is simply the coplanar 
AFM$_1$ state in Fig.~\ref{fig5}. It is thus of interest to search for topological 
magnons in this material. Tl$_2$Ru$_2$O$_7$ experiences a structural transition 
at 120K that breaks the cubic symmetry, so our model does not really apply here. 
Eu$_2$Ru$_2$O$_7$ was suggested to develop Ru sublattice orders at 118K
and experience a glassy-like transtion at 23K~\cite{Eu2Ru2O7}. The precise 
nature of the Ru order is not known. 

The Ru materials with the unquenched rare-earth moments contain richer physics than
the ones with non-magnetic rare-earth moments. There are three energy scales to consider.
From high to low in the energy scales, we would list them as Ru-Ru exchange interaction,
$f$-$d$ exchange between the Ru moments and rare-earth moments, and the 
exchange and dipolar interactions between the rare-earth moments. This hierarchical
energy structure arises from the different spatial extension of the $4d$ electrons
and the $4f$ electrons. Since the Ru-Ru exchange interaction
would be the dominant one, we would expect the Ru moments to develop structures at 
higher temperatures and influence the rare-earth moments via the $f$-$d$ exchange. 
The existing experiments support this view~\cite{RevModPhys.82.53}.   

The experimental study on these rare-earth based Ru pyrochlores has not been quite systematic yet.
Only limited experimental information is available. We here focus the discussion on
the systems with more known results. Ho$_2$Ru$_2$O$_7$ was studied using neutron scattering 
measurements in a nice paper~\cite{PhysRevLett.93.076403} by C.R. Wiebe, {\it et al}. 
The authors revealed the Ru moment order at ${\sim95}$K and the Ho moment order at ${\sim1.4}$K. 
The high temperature Ru magnetic order is consistent with the splayed FM with a splayed angle
${\alpha} \approx 41$ degrees. Under the internal exchange field from the Ru order, 
the Ho moment further develops a magnetic order at a lower temperature. Despite the agreement
between the experimental order and theoretical result, further measurement of the magnetic 
excitation within the splayed FM can be useful to identify nontrival magnon modes. 
Ref.~\onlinecite{ErRuO} carried out a powder neutron scattering measurement 
on Er$_2$Ru$_2$O$_7$ and proposed a ${{\boldsymbol Q}={\boldsymbol 0}}$ 
with a collinear antiferromagnetic magnetic order along ${\langle001\rangle}$ lattice 
direction for the Ru moments. Like the Ho$_2$Ru$_2$O$_7$, the Er moments develop a 
magnetic order at a much lower temperature. Since the Ru moment ordering occurs 
at much higher temperature and should be understood first. To stabilize the 
collinear order for the Ru moments, one may need a biquadratic 
spin interaction~\cite{PhysRevB.74.134409,PhysRevLett.93.197203}.
This collinear state is actually not among the ordered states that we find. 
We suscept one ordered state in Fig.~\ref{fig5} may also explain the existing 
data for Er$_2$Ru$_2$O$_7$. More experiments are needed to sort out the actual 
magnetic order in this material.  
 
Because the Ru spin-1 moments in these materials often order at a higher temperature, 
it would be interesting to examine the precise magnetic structure and the magnetic 
excitations in the future experiments and compare with the theoretical prediction. 
Future theoretical directions in these systems at least include the understanding 
of the $f$-$d$ exchange between the rare-earth moments and the Ru moments and the 
magnetic properties of the rare-earth subsystem. The $f$-$d$ exchange significantly 
depends on the nature of the rare-earth moment, i.e. whether it is Kramers doublet, 
non-Kramers doublet or dipole-octupole doublet. As a result, the Ru molecular or 
internal exchange field on the rare-earth subsystem not only depends on the magnetic 
structure of the Ru subsystem, but also depends on the form of the $f$-$d$ exchange. 
This may give rise to rich magnetic structures and properties on the rare-earth
subsystems in the ordered phase of the Ru subsystems.  

\begin{figure}[b]
	\includegraphics[width=7cm]{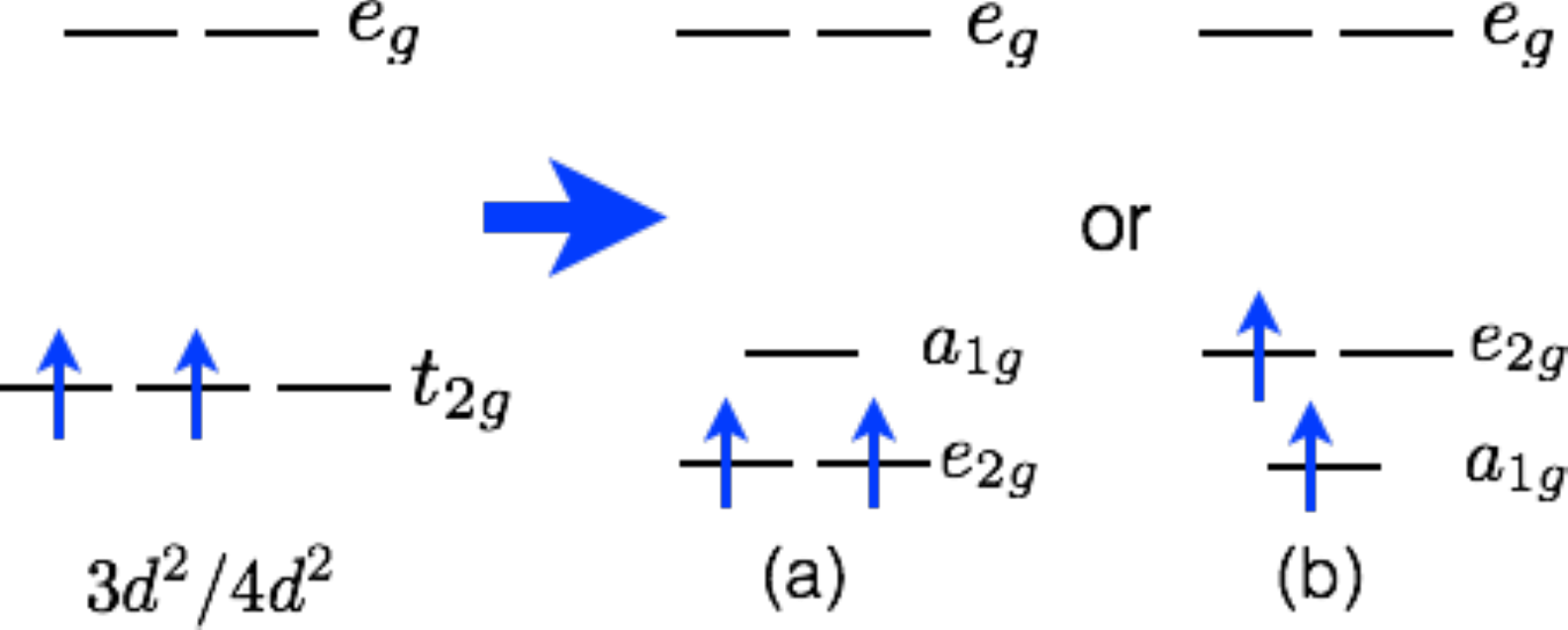}
	\caption{The orbital occupations for $3d^2/4d^2$ electron configuration. 
	Under the trigonal distortion, the three-fold degenerate $t_{2g}$
	orbitals are splitted into $a_{1g}$ and two-fold degenerate $e_{2g}$ states. 
	There are two electron occupation configurations here. (b) has an unquenched orbital degree
	of freedom.}
	\label{figx2}
\end{figure}

It is interesting to compare the spin-1 Ru pyrochlores with the rare-earth 
osmates (A$_2$Os$_2$O$_7$) and molybedates (A$_2$Mo$_2$O$_7$). 
The Os$^{4+}$ ion has a $5d^4$ electron configuration, 
and spin-orbit coupling is stronger than Ru$^{4+}$. As a result, rather 
than forming a ${S=1}$ local moment, the magnetic moment of the Os$^{4+}$ 
ion is strongly suppressed by the spin-orbit coupling that would 
favor a spin-orbital singlet in the strong spin-orbit coupling 
limit~\cite{PhysRevB.84.094420,PhysRevB.93.134426,PhysRevLett.111.197201}.  
Unlike the insulating Ru-based pyrochlores, most Mo-based pyrochlore materials 
are metallic~\cite{RevModPhys.82.53}. The Mo$^{4+}$ has a $4d^2$ electron 
configuration. The metallic behavior is probably because the Hund's coupling 
suppresses the correlation effect and induces Hund's metals~\cite{hundsmetal}.
Instead of developing magnetic orders, the insulating ones (Y$_2$Mo$_2$O$_7$,
Lu$_2$Mo$_2$O$_7$ and Tb$_2$Mo$_2$O$_7$) all show spin glassy behaviors. 
The origin of the spin glass in these geometrically frustrated 
pyrochlore molybedates remains to be a puzzle in the field~\cite{RevModPhys.82.53}. 
It is possible that, the orbital occupation of the Mo$^{4+}$ ion is not given
by Fig.~\ref{figx2}a and is instead given by Fig.~\ref{figx2}b. In that case,
the Mo local moment contains a unquenched orbital degree of freedom, and the orbital and 
spin interact in a Kugel-Khomskii fashion~\cite{Kugel82} and are affected by the lattice 
phonons. This spin-orbital physics has been suggested for the spinel vanadate AV$_2$O$_4$ 
(A=Ca,Mg,Cd,Zn), where ${V^{3+}: 3d^2}$ was expected to take the 
electron configuration in Fig.~\ref{figx2}b ~\cite{PhysRevLett.93.156407,PhysRevLett.99.126401,PhysRevB.83.060402,
PhysRevLett.94.156402,PhysRevLett.111.267201,PhysRevB.82.140406} and 
forms a spin-1 pyrochlore system with additional orbital degree of freedom. 

\subsection{Extension to spin-3/2 pyrochlores}

Although the major part of the paper deals with the spin-1 pyrochlore materials, 
the same model actually applies to the spin-3/2 pyrochlore materials. 
The spin-3/2 moment is a half-integer moment, and the local spin anisotropy
acts on it quite differently from the spin-1 moment. Certainly the quantum 
paramagnetic phase is absent for the spin-3/2 moment, and there are always 
unquenched local moments regardless of the easy-axis or easy-plane anisotropy. 
In the strong easy-axis or easy-plane anisotropic limit, the spin-3/2 moment   
reduces to an effective spin-1/2 moment that can be described by the generic 
and anisotropic model for the effective spin-1/2 moment. This point of view  
has been suggested for the Co-based pyrochlore materials NaCaCo$_2$F$_7$ and 
NaSrCo$_2$F$_7$ in Ref.~\onlinecite{PhysRevB.95.144414}. 
Besides this difference from the spin-1 moment, the magnetic orders, if 
they occur in the spin-3/2 pyrochlore system, would be similar to the spin-1 
pyrochlore system, since the same mean-field Hamiltonian applies to both 
systems. Moreover, the spin wave excitation would have similar properties. 
For example, we would expect the existence of the topological spin wave modes
such as Weyl magnons in the magnetic excitations of the ordered spin-3/2 
pyrochlore materials. In fact, the notion of Weyl magnon was first proposed 
by our collaborators and us in the context of the Cr-based spin-3/2 breathing 
pyrochlore systems. The model Hamiltonian, that was used in Ref.~\onlinecite{Weylmagnon} 
did not include the Dzyaloshinskii-Moriya interaction. It was also shown 
in Ref.~\onlinecite{Weylmagnon} that the Weyl magnon is robust against 
weak perturbation and extends to the regime of a regular pyrochlore system. 
Besides the Co-pyrochlore and Cr-spinel, the Mn-pyrochlore (A$_2$Mn$_2$O$_7$)
is another ideal spin-3/2 system. These materials were studied in the 90s 
after the discovery of giant magnetoresistance~\cite{RevModPhys.82.53}.  
Since most of these Mn-pyrochlores are well ordered, it would be exciting 
to explore the topological magnons in these materials.

\section{Acknowledgments}

We are indebted to D.-H. Lee and F.-C. Zhang for their advice that 
wakes me up to write out and/or submit our papers including this one here. 
We thank T. Senthil for a conversation at the Hong Kong Gordon Research 
Conference this June about the pyrochlore lattice Heisenberg model 
on which we shared some common view and have actually expressed in 
Ref.~\onlinecite{GangChen2017} before our conversation. This work 
is supported by the ministry of science and technology of China with 
the grant No.2016YFA0301001, the start-up fund and the first-class 
University construction fund of Fudan University, and the 
thousand-youth-talent program of China.

\appendix

\section{Dzyaloshinskii-Moriya interaction}
\label{app1}

Below we list ${\boldsymbol D}_{ij}$ vectors in the Dzyaloshinskii-Moriya
interaction~\cite{PhysRevB.71.094420} for bonds in Fig.~\ref{fig2}:
\begin{eqnarray}
{\boldsymbol  D}_{01}&=&\frac{1}{\sqrt{2}}(0,+D,-D) ,   \\
{\boldsymbol  D}_{02}&=&\frac{1}{\sqrt{2}}(-D,0,+D) ,   \\
{\boldsymbol  D}_{03}&=&\frac{1}{\sqrt{2}}(+D,-D, 0) ,  \\
{\boldsymbol  D}_{12}&=&\frac{1}{\sqrt{2}}(+D,+D,0),	   \\
{\boldsymbol  D}_{13}&=&\frac{1}{\sqrt{2}}(-D,0,-D),    \\
{\boldsymbol  D}_{23}&=&\frac{1}{\sqrt{2}}(0,+D,+D).
\end{eqnarray}

\section{Transformation of the model}
\label{app2}

We first define the local coordinate systems where $S^z_i$ and $S^{\pm}_i$
are defined. The choices of the local spin axes are listed in Table~\ref{stab1}. 

\begin{table}[ht]
\begin{tabular}{lcccc}
\hline\hline
${\mu}$          & 0 & 1 & 2 & 3 
\\[.2cm]
$\hat{x}_{\mu}$  & $\frac{1}{\sqrt{6}}[\bar{2}11]$ & $\frac{1}{\sqrt{6}} [\bar{2}\bar{1}\bar{1}]$ 
                 & $\frac{1}{\sqrt{6}}[21\bar{1}]$ & $\frac{1}{\sqrt{6}} [2\bar{1}1] $ 
\\[.2cm]
$\hat{y}_{\mu}$  & $\frac{1}{\sqrt{2}}[0\bar{1}1]$ & $\frac{1}{\sqrt{2}} [01\bar{1}]$ 
                 & $\frac{1}{\sqrt{2}}[0\bar{1}\bar{1}]$ & $\frac{1}{\sqrt{2}}[011]$
\\[.2cm]
$\hat{z}_{\mu}$  & $\frac{1}{\sqrt{3}}[111]$ &  $\frac{1}{\sqrt{3}} [1\bar{1}\bar{1}]$ 
                 & $\frac{1}{\sqrt{3}} [\bar{1}1\bar{1}]$ & $\frac{1}{\sqrt{3}}[\bar{1}\bar{1}1]$
\\[.2cm]
\hline\hline
\end{tabular}
\caption{The local coordinate systems for the four sublattices. The same choice 
can be found for the spin-1/2 Kramers doublet in Ref.~\onlinecite{Ross11}.} 
\label{stab1}
\end{table}

The relation between the couplings in Eq.~\eqref{eq1}
and the couplings in Eq.~\eqref{eq4} is given as
\begin{eqnarray}
J_{zz} &=& \frac{1}{3}\left(2\sqrt{2}D-J\right), \nonumber \\
J_{\pm} &=& -\frac{1}{6}\left(\sqrt{2}D+J\right), \nonumber \\
J_{\pm\pm} &=& -\frac{1}{3}\left(\frac{D}{\sqrt{2}}-J\right), \nonumber \\
J_{z\pm} &=& \frac{1}{6}\left(D+2\sqrt{2}J\right).
\end{eqnarray}

The bond-dependent phase variables $\gamma_{ij}$ and $\xi_{ij}$ 
can be written in matrix form as
\begin{equation}
\gamma=-\xi^*=
\begin{pmatrix}
0 & 1 & \text{e}^{i2\pi/3} & \text{e}^{-i2\pi/3}
\\
1 & 0 & \text{e}^{-i2\pi/3} & \text{e}^{i2\pi/3}
\\
\text{e}^{i2\pi/3} & \text{e}^{-i2\pi/3} & 0 & 1
\\
\text{e}^{-i2\pi/3} & \text{e}^{i2\pi/3} & 1 & 0
\end{pmatrix},
\end{equation}
where the indices of the matrix label different sublattices.

\section{Flavor wave Hamiltonian}
\label{app3}

The flavor wave Hamiltonian matrix defined in Eq.~\eqref{eq12}
can be written in block form as
\begin{equation}
M(\boldsymbol{k})=
\begin{pmatrix}
M_1(\boldsymbol{k}) & M_2(\boldsymbol{k})
\\
M_2^*(\boldsymbol{k}) & M_1^*(\boldsymbol{k})
\end{pmatrix},
\end{equation}
where $M_1(\boldsymbol{k})$ and $M_2(\boldsymbol{k})$ are $8\times8$ 
matrices and satisfy $M_1^\dagger(\boldsymbol{k})=M_1(\boldsymbol{k})$, 
$M_2^T(\boldsymbol{k})=M_2(\boldsymbol{k})$.

$M_1(\boldsymbol{k})$ and $M_2(\boldsymbol{k})$ can be further written 
in block form as
\begin{equation}
\begin{pmatrix}
m_{00} & m_{01} & m_{02} & m_{03}
\\
m_{10} & m_{11} & m_{12} & m_{13}
\\
m_{20} & m_{21} & m_{22} & m_{23}
\\
m_{30} & m_{31} & m_{32} & m_{33}
\end{pmatrix},
\end{equation}
where $m_{\mu\nu}$s are $2\times2$ matrices.

\begin{figure}[t]
	\includegraphics[width=8.5cm]{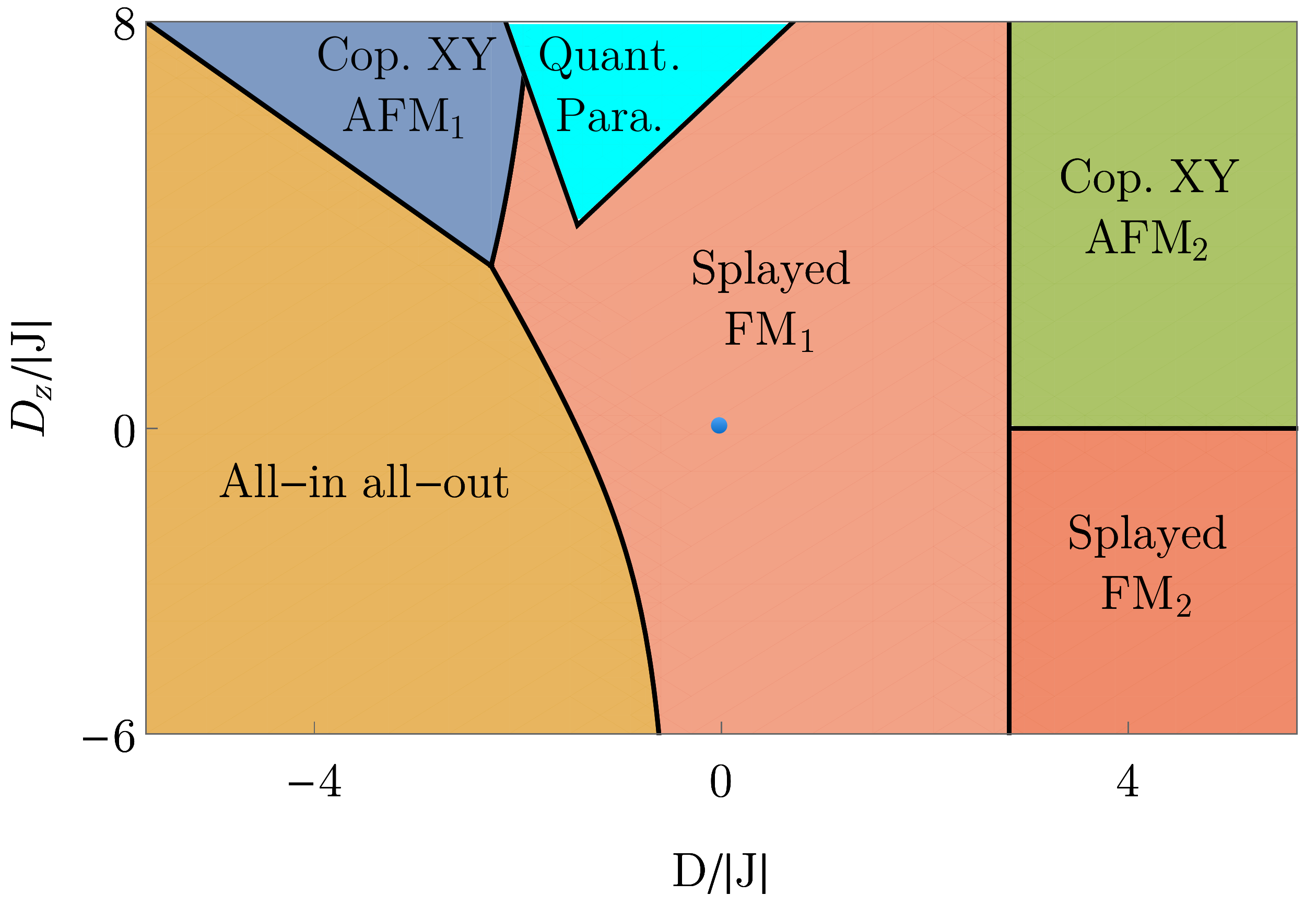}
	\caption{
		The phase diagram of our generic spin model with 
		ferromagnetic Heisenberg exchange (${J<0}$). The 
		Heisenberg point is labeled by a blue point.}
	\label{fig12}
\end{figure}

For $M_1(\boldsymbol{k})$,
\begin{eqnarray}
m_{\mu\mu}&=&\frac{1}{2}
\begin{pmatrix}
D_z & 0
\\
0 & D_z
\end{pmatrix}, \nonumber \\
m_{\mu\nu(\mu\neq\nu)}&=&
2\cos\Phi_{\mu\nu}
\begin{pmatrix}
J_{\pm} & J_{\pm\pm}\gamma_{\mu\nu}
\\
J_{\pm\pm}\gamma^*_{\mu\nu} & J_{\pm}
\end{pmatrix}.
\end{eqnarray}
For convenience, here and henceforth we define
\begin{equation}
\Phi_{\mu\nu}\equiv\boldsymbol{k}\cdot\left({\boldsymbol{r}_\mu}-{\boldsymbol{r}_\nu}\right) ,
\end{equation} 
where
$\boldsymbol{r}_{0}=[000]$, $\boldsymbol{r}_{1}=\frac{1}{4}[011]$, 
$\boldsymbol{r}_{2}=\frac{1}{4}[101]$, $\boldsymbol{r}_{3}=\frac{1}{4}[110]$.

For $M_2(\boldsymbol{k})$,
\begin{eqnarray}
m_{\mu\mu}&=&
\begin{pmatrix}
0 & 0
\\
0 & 0
\end{pmatrix}, \nonumber \\
m_{\mu\nu(\mu\neq\nu)}&=&
2\cos\Phi_{\mu\nu}
\begin{pmatrix}
J_{\pm\pm}\gamma_{\mu\nu} & J_{\pm}
\\
J_{\pm} & J_{\pm\pm}\gamma^*_{\mu\nu}
\end{pmatrix}.
\end{eqnarray}

\section{Spin wave Hamiltonian} 
\label{app4}

\begin{figure}[b]
\includegraphics[width=6cm]{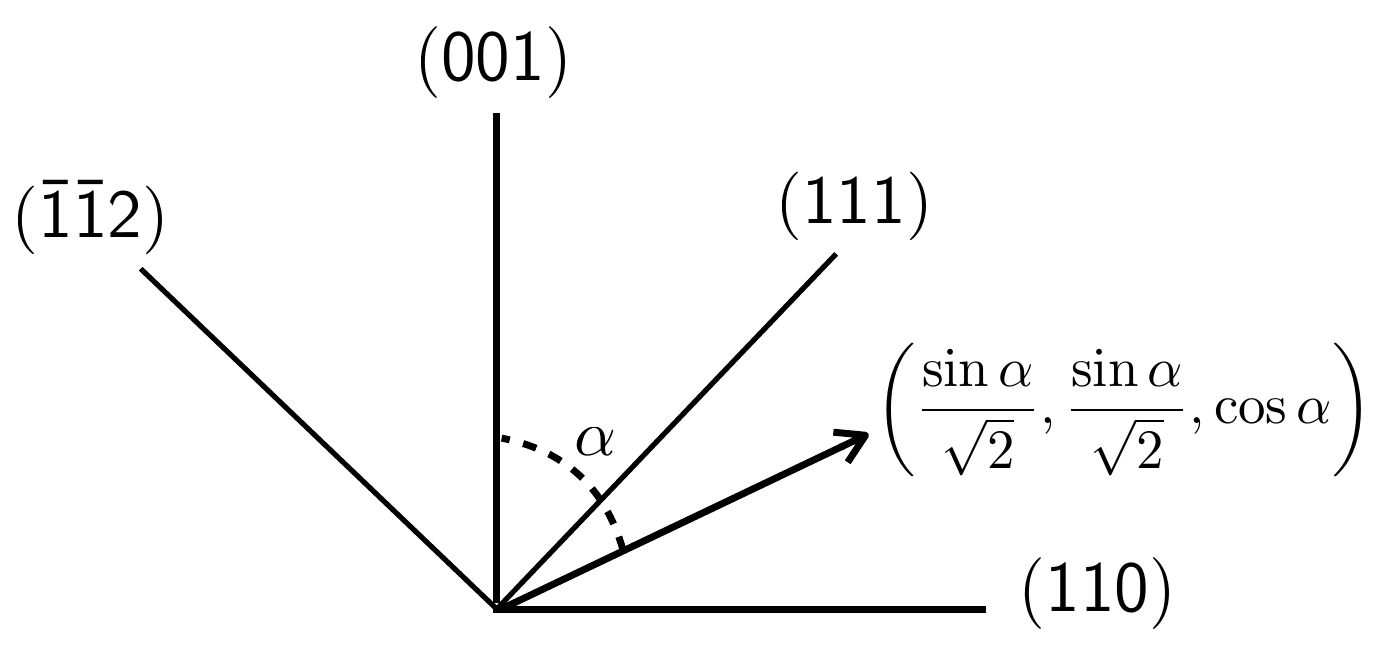}
\caption{The parameter regime of the splay angle $\alpha$ in 
``splayed FM$_1$'' and ``splayed FM$_2$''. For the splayed ferromagnet, 
the spin on sublattice 0 can be parametrized as  $\left({\sin\alpha}/{\sqrt{2}}, 
{\sin\alpha}/{\sqrt{2}}, \cos\alpha\right)$, where we set the ferromagnetic 
component along $z$ direction. In ``splayed FM$_1$'' and for fixed $D$, 
when $D_z$ is tuned from negative infinity to positive infinity, the spin 
on sublattice 0 sweeps from (111) to ($\bar{1}\bar{1}2$) and $\alpha$ takes 
value from $54.7^\circ$ to $-35.3^\circ$. When $\alpha=0$, we have a collinear 
ferromagnetic state. On the other hand, in ``splayed FM$_2$'' and for fixed $D$, 
the spin on sublattice 0 sweeps from (110) to (111) when $D_z$ is tuned away 
from 0. The splay angle $\alpha$ then takes value from 90$^\circ$ to 54.7$^\circ$. 
When $\alpha=90^\circ$, we have a coplanar state.}
	\label{fig13}
\end{figure}

The entries of the spin wave Hamiltonian in Eq.~\eqref{eq27} are given as
\begin{eqnarray}
&&A_{\mu\mu}=2\sqrt{2}D+D_z+2J,  \nonumber \\
&&A_{\mu\nu(\mu\neq\nu)}=\frac{\cos \Phi_{\mu\nu}}{3}
(\sqrt{2}D-2J)\Big[1+\cos(2\theta+\phi_{\mu\nu})\Big],  \nonumber \\
&&B_{\mu\mu}=\frac{1}{2}D_z, \nonumber \\
&&B_{\mu\nu(\mu\neq\nu)}=-\frac{1}{6} \cos \Phi_{\mu\nu}
\Big[   (\sqrt{2}D-2J)\cos(2\theta+\phi_{\mu\nu}) \nonumber  \\
&&\quad\quad \quad\quad  +i(2D+4\sqrt{2}J)\sin(\theta-\phi_{\mu\nu})-3\sqrt{2}D\Big], 
\nonumber
\end{eqnarray}
where the angle variable $\phi_{\mu\nu}$ is given as
\begin{equation}
\phi=
\begin{pmatrix}
0 & 0 & 2\pi/3 & -2\pi/3
\\
0 & 0 & -2\pi/3 & 2\pi/3
\\
2\pi/3 & -2\pi/3 & 0 & 0
\\
-2\pi/3 & 2\pi/3 & 0 & 0
\end{pmatrix}.
\end{equation}

\section{Ferromagnetic phase diagram}
\label{app5}

In Fig.~\ref{fig12}, we show the ferromagnetic phase diagram of our 
generic spin model defined in Eq.~\eqref{eq1}. In the phase diagram, 
``quant para'' refers to the quantum paramagnetic phase and the 
other regions are ordered phases. ``All-in all-out'', ``coplanar 
XY AFM$_1$'' and ``coplanar XY AFM$_2$''  are the same phases as 
described in the antiferromagnetic phase diagram of Fig.~\ref{fig1}. 
The splayed ferromagnet is divided into ``splayed FM$_1$'' and 
``splayed FM$_2$'' according to the parameter regime of the splay 
angle $\alpha$, demonstrated in Fig.~\ref{fig13}.

\section{Order selection on the phase boundaries}
\label{app6}

\subsubsection{${D_z=\sqrt{2}|D|}$}

On the line ${D_z=\sqrt{2}|D|}$, one have three sets of the ground states with $U(1)$ 
degeneracy. Combining the ``all-in all-out'' configuration and the configuration in 
Eq.~\eqref{eq20}, one set of the ground states can be parametrized as
\begin{eqnarray}
\left\{ \begin{array}{l}
{\boldsymbol m}_0  =  \cos\varphi~ \frac{1}{\sqrt{3}}(1,1,1) 
                    + \sin\varphi~ \frac{1}{\sqrt{2}} (1, \bar{1}, 0) ,
\\[.4cm]
{\boldsymbol m}_1  =  \cos\varphi~ \frac{1}{\sqrt{3}}(1,\bar{1},\bar{1}) 
                    + \sin\varphi~ \frac{1}{\sqrt{2}} (\bar{1}, \bar{1}, 0) ,
\\[.4cm]
{\boldsymbol m}_2  =  \cos\varphi~ \frac{1}{\sqrt{3}}(\bar{1},1,\bar{1}) 
                    + \sin\varphi~ \frac{1}{\sqrt{2}} (1,{1},{0}) ,
\\[.4cm]
{\boldsymbol m}_3  =  \cos\varphi~ \frac{1}{\sqrt{3}}(\bar{1},\bar{1},1) 
                    + \sin\varphi~ \frac{1}{\sqrt{2}} (\bar{1}, {1},0) .
\end{array}
\right.
\label{eq-a}
\end{eqnarray}
The other two sets are symmetry equivalent and can be obtained by a three-fold rotation.

\begin{figure}[t]
\includegraphics[width=7.8cm]{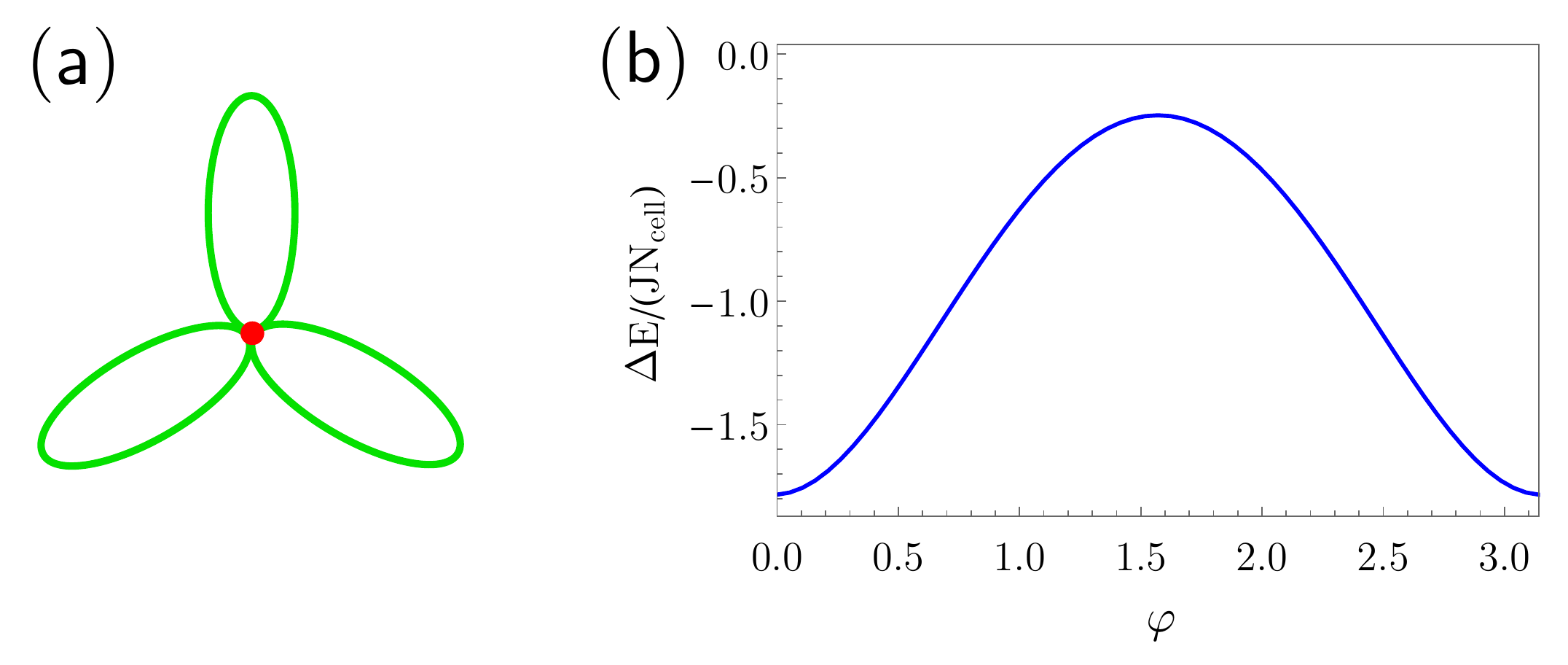}
\caption{Order by quantum disorder on the line ${D_z=\sqrt{2}|D|}$. 
(a) Three ellipses represent three sets of the ground states with $U(1)$ degeneracy.
Their intersection point corresponds to the ``all-in all-out'' state.
(b) For the parametrization in Eq.~\eqref{eq-a}, the minima of the quantum 
zero-point energy are realized at ${\varphi=0,\pi}$, selecting the ``all-in all-out'' 
state (indicated by a red point). The parameters are ${D_z=\sqrt{2}|D|=0.5J}$.
Here and in Figs.~\ref{fig9},~\ref{fig10}, we adopt the Fig.~7 of 
Ref.~\onlinecite{PhysRevB.71.094420} to schematically represent the ground state 
manifold and the order by quantum disorder effect. Note that each state and its 
time reversal partner are represented by a single point on the ellipses or circles.}
\label{fig7}
\end{figure}

For each set of the ground states, the minima of the quantum zero-point energy are realized 
at ${\varphi=0,\pi}$, so the order by quantum disorder effect selects the ``all-in all-out'' 
state, see Fig.~\ref{fig7}.

\subsubsection{${D=0,D_z>0}$}

When Dzyaloshinskii-Moriya interaction is switched off, the model 
describes an anisotropic pyrochlore lattice antiferromagnet.
Although the easy-axis anisotropy (${D_z<0}$) leads to 
simple "all-in all-out" configuration in mean-field level,
the easy-plane case (${D_z>0}$) has a rich structure of 
the ground state manifold.

First, we have a $U(1)$ ground state manifold defined as
\begin{eqnarray}
{\boldsymbol m}_{\mu} =  \hat{x}_{\mu} \cos \theta 
                       + \hat{y}_{\mu} \sin \theta.
\end{eqnarray}
For convenience we now dub this manifold ``XY$_0$''. 
Combining XY$_0$ and the ground state configurations of ``coplanar XY AFM$_1$'',
one can construct extra three sets of generally non-coplanar XY AFM ground states 
with $U(1)$ degeneracy, dubbed ``XY$_1$'',  ``XY$_2$'' and  ``XY$_3$'', respectively.

We here define the local direction ${\hat{n}_{\mu}^{\phi}\equiv \hat{x}_{\mu} \cos \phi 
+ \hat{y}_{\mu} \sin \phi}$, where $\phi$ is a rotation angle in the local $xy$ plane.
The XY$_1$ ground states are parametrized as
\begin{eqnarray}
\left\{ \begin{array}{l}
{\boldsymbol m}_0  =  \cos\varphi ~\hat{n}_{0}^{{\frac{\pi}{3}}} 
  + \sin\varphi ~\frac{1}{\sqrt{2}} (1, \bar{1}, 0) ,
\\[.4cm]
{\boldsymbol m}_1  =  \cos\varphi ~\hat{n}_{1}^{{\frac{\pi}{3}}} 
  + \sin\varphi ~\frac{1}{\sqrt{2}} (\bar{1}, \bar{1}, 0) ,
\\[.4cm]
{\boldsymbol m}_2  =  \cos\varphi ~\hat{n}_{2}^{{\frac{\pi}{3}}} 
  + \sin\varphi ~\frac{1}{\sqrt{2}} (1,{1},{0}) ,
\\[.4cm]
{\boldsymbol m}_3  =  \cos\varphi ~\hat{n}_{3}^{{\frac{\pi}{3}}} 
  + \sin\varphi ~ \frac{1}{\sqrt{2}} (\bar{1}, {1},0) .
\end{array}
\right.
\label{eq-b}
\end{eqnarray}
The symmetry related XY$_2$ and XY$_3$ ground states can be obtained 
by applying the space group symmetry operations.

\begin{figure}[b]
	\includegraphics[width=6cm]{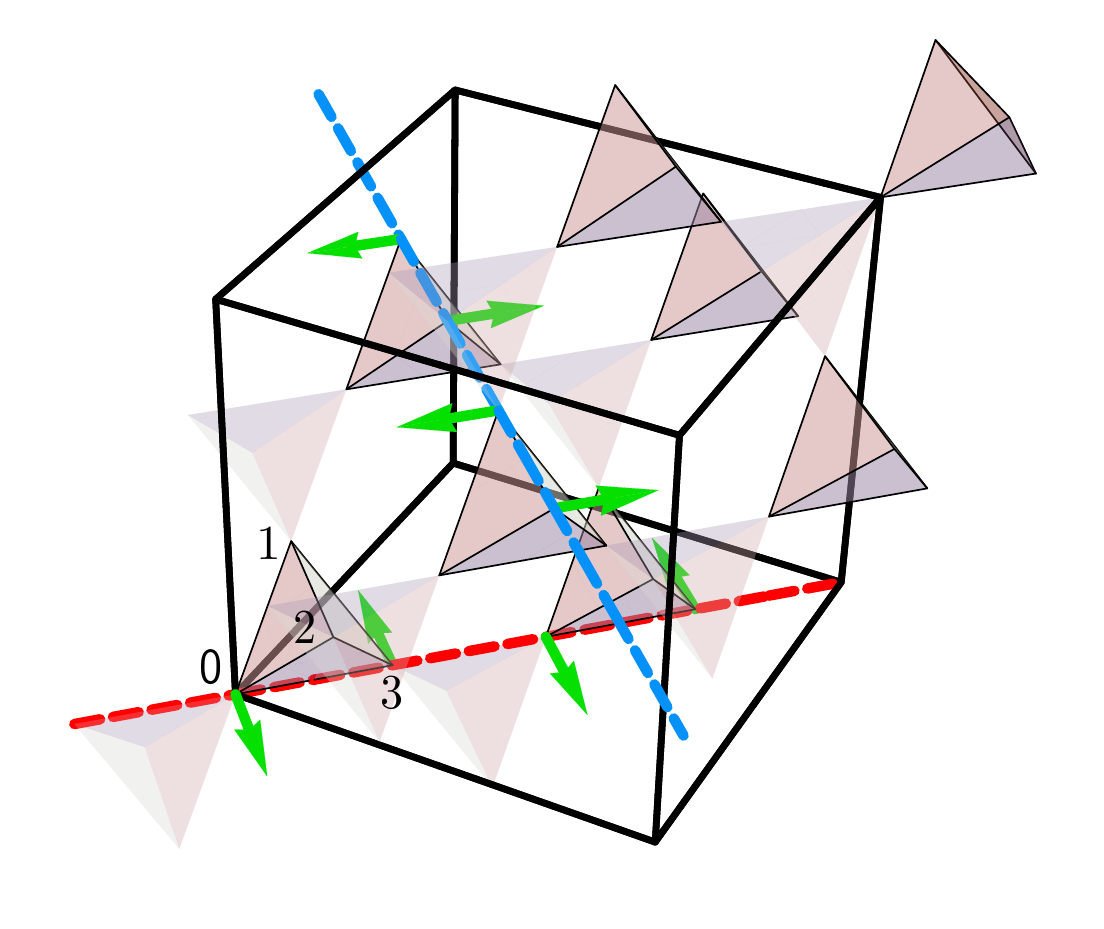}
	\caption{The way to construct ground states with discrete degeneracy. 
		For clarity, we only depict the part of the spin configuration 
		given in Eq.~\eqref{eq20} (green arrows).
		Start from this state, the freedom of simultaneously flipping the 
		spins along any 0-3-0-3-$\cdots$ chain (red dashed line) 
		or 1-2-1-2-$\cdots$ chain (blue dashed line) leads to hugely degenerate ground states.}
	\label{fig8}
\end{figure}

Moreover, one can construct ground states with huge discrete 
degeneracy~\cite{J.Appl.Phys.75.5523}.
This can be understood like this~\cite{J.Appl.Phys.75.5523}: 
to optimize the antiferromagnetic Heisenberg interaction, one needs
to arrange $\sum_\mu {\boldsymbol m}_\mu=0$ in each tetrahedron, and 
to satisfy $D_z$ term ${\boldsymbol m}_\mu$, the spins must orient within 
the local $xy$ plane. Starting from the state defined in Eq.~\eqref{eq20} 
where for this state ${{\boldsymbol m}_0+{\boldsymbol m}_3=0}$ and 
${{\boldsymbol m}_1+{\boldsymbol m}_2=0}$ are satisfied in each tetrahedron
and each spin orients within the local $xy$ plane, we can simultaneously 
flip the spins along any 0-3-0-3-$\cdots$ chain or 1-2-1-2-$\cdots$ chain 
without changing the mean-field energy (see Fig.~\ref{fig8}). 
Repeating this process, one obtains $4^{N^{2/3}}$ 
degenerate states where $N$ is the total number of the unit cells. 
These states are coplanar states in the global $xy$ plane and generally 
have no translational symmetry. Similar coplanar states in the global 
$yz$ and $zx$ plane can be readily obtained by applying a three-fold 
rotation.

\begin{figure}[h]
	\includegraphics[width=7.8cm]{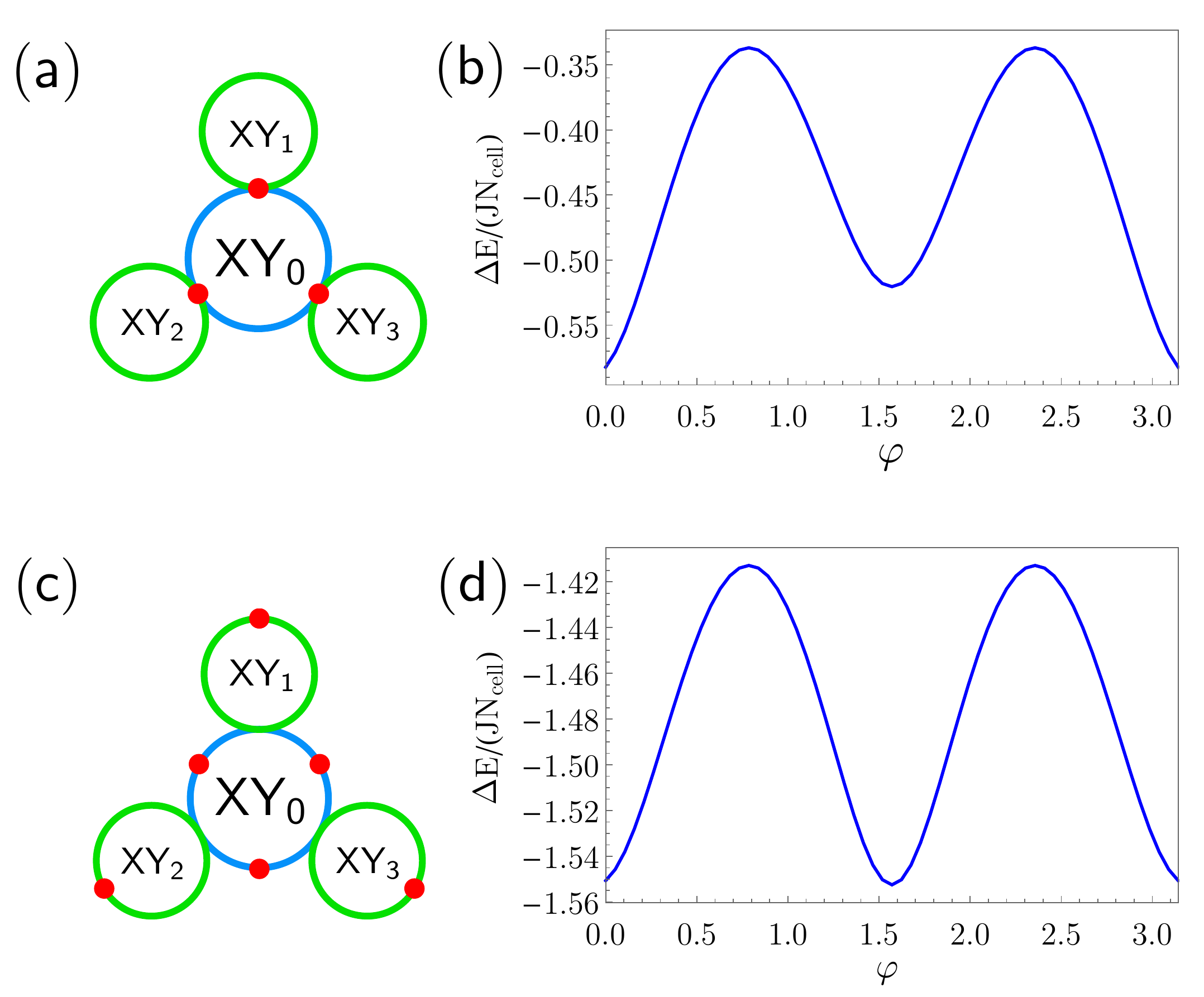}
	\caption{Order by quantum disorder on the line ${D=0,D_z>0}$. 
		(a)[(c)], four circles represent four sets of ground states with $U(1)$ degeneracy and
		the red points indicate the states selected by quantum fluctuation for ${D_z>0.11J}$
		(${D_z<0.11J}$).
		For the parametrization in Eq.\eqref{eq-b}, the minima of the quantum zero-point 
		energy are realized at (b) ${\varphi=0,\pi}$ with ${D=0,D_z=0.5J}$;
		(d) $\varphi={\pi}/{2},{3\pi}/{2}$ with ${D=0,D_z=0.1J}$.
	}
	\label{fig9}
\end{figure}

Now we discuss the order by quantum disorder effect for the 
 ground state manifold with continuous degeneracy, 
There is a boundary point ${D_z=0.11J}$ separating the 
``non-coplanar XY AFM'' and the ``coplanar XY AFM$_2$''
along the ${D=0}$ line, and the order by quantum disorder effect 
naturally depends on $D_z$. For ${D_z>0.11J}$, the minima of 
the quantum zero-point energy select the ground states of 
``non-coplanar XY AFM'' from the continuous manifold, 
see Fig.~\ref{fig9}(a)(b). For ${D_z<0.11J}$, the ground states of 
``coplanar XY AFM$_1$'' and ``coplanar XY AFM$_2$''  are selected 
ground states when quantum fluctuation is included (see Fig.~\ref{fig9}(c)(d)).

We mention that all the mean-field ground states found here still hold as 
ground states for an anisotropic antiferromagnetic Heisenberg model on 
the breathing pyrochlore lattice, which is previously discussed in 
Ref.~\onlinecite{Weylmagnon}.

\subsubsection{${D_z=0,D>0}$}

When the anisotropy is absent, a negative Dzyaloshinskii-Moriya interaction  
favors simple ``all-in all-out'' state, and a positive Dzyaloshinskii-Moriya 
interaction leads to a ground state manifold with continuous degeneracy. 
This regime has been studied in the previous work by mean-field theory
and classical Monte carlo~\cite{PhysRevB.71.094420}. We here explore the 
quantum effect beyond the mean-field theory. 

Besides the XY$_0$ manifold, we have another three sets of coplanar 
ground states in the case of a positive Dzyaloshinskii-Moriya interaction. 
The ``splayed FM'' states become coplanar when approaching the limit $D_z=0$. 
One such state is
\begin{eqnarray}
\left\{ \begin{array}{l}
{\boldsymbol m}_0  =  \frac{1}{\sqrt{2}}(1,1,0) ,
\\[.4cm]
{\boldsymbol m}_1  =  \frac{1}{\sqrt{2}}(\bar{1},1,0) ,
\\[.4cm]
{\boldsymbol m}_2  =  \frac{1}{\sqrt{2}}(1,\bar{1},0) ,
\\[.4cm]
{\boldsymbol m}_3  =  \frac{1}{\sqrt{2}}(\bar{1},\bar{1},0).
\end{array}
\right.
\label{eq-c}
\end{eqnarray}

Combining this state with proper state in the XY$_0$ manifold, 
one can construct a set of coplanar ground states in the global 
$xy$ plane, parametrized as
\begin{eqnarray}
\left\{ \begin{array}{l}
{\boldsymbol m}_0  =  \cos\varphi ~\hat{n}_{0}^{{-\frac{\pi}{6}}} + \sin\varphi ~\frac{1}{\sqrt{2}}(1,1,0) ,
\\[.4cm]
{\boldsymbol m}_1  =  \cos\varphi ~\hat{n}_{1}^{{-\frac{\pi}{6}}} + \sin\varphi ~\frac{1}{\sqrt{2}}(\bar{1},1,0) ,
\\[.4cm]
{\boldsymbol m}_2  =  \cos\varphi ~\hat{n}_{2}^{{-\frac{\pi}{6}}} + \sin\varphi ~\frac{1}{\sqrt{2}}(1,\bar{1},0) ,
\\[.4cm]
{\boldsymbol m}_3  =  \cos\varphi ~\hat{n}_{3}^{{-\frac{\pi}{6}}} + \sin\varphi ~\frac{1}{\sqrt{2}}(\bar{1},\bar{1},0) .
\end{array}
\right.
\end{eqnarray}
Again the other two sets of coplanar ground states, 
in the global $yz$ and $zx$ plane respectively, can 
be obtained by applying the three-fold rotation.

\begin{figure}[h]
\includegraphics[width=7.8cm]{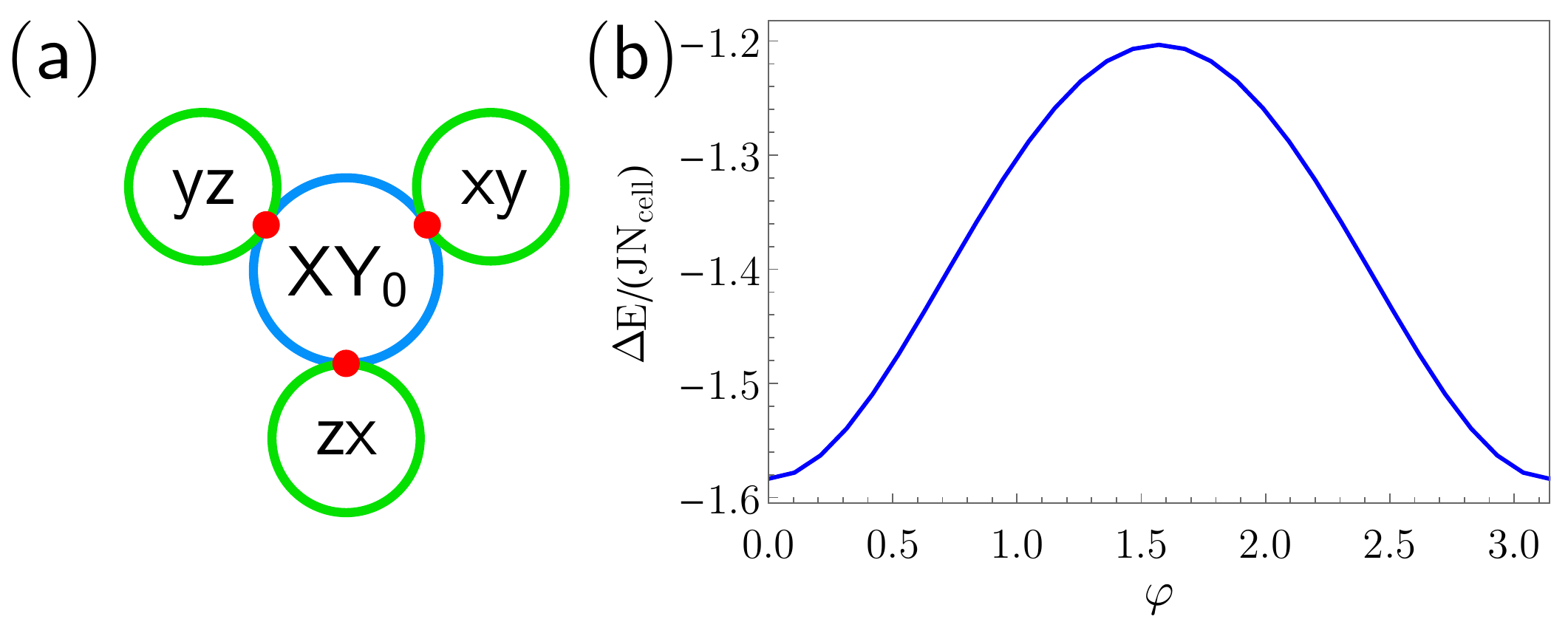}
\caption{Order by quantum disorder on the line $D_z=0, D>0$.
(a) Four circles represent four sets of ground states with $U(1)$ degeneracy 
and the red points indicate the states selected by quantum fluctuation. 
We refer three sets of coplanar states as ``xy'', ``yz'', ''zx'' respectively.
(b) For the parametrization in Eq.~\ref{eq-c}, the minima of zero-point energy 
are realized at $\varphi=0,\pi$ with $D_z=0, D=0.5J$.
}
\label{fig10}
\end{figure}

When one includes quantum fluctuation, it turns out that the minima of the 
quantum zero-point energy select the ground states of ``coplanar XY AFM$_1$'' 
from the whole manifold, see Fig.~\ref{fig10}. 

The ground state structure of the line ${D_z=0,D>0}$ and the order by disorder effect 
(quantum and thermal) have been extensively studied~\cite{PhysRevB.71.094420,PhysRevB.78.214431,1008.3038}.
We mention that it is more natural to understand the four-set structure of the ground 
state manifold by putting this line on the full phase diagram in Fig.~\ref{fig1}.

\bibliography{refs}

\end{document}